\documentclass[aps,prb,twocolumn,longbibliography,superscriptaddress]{revtex4-2}
\pdfoutput=1

\usepackage{amsmath}
\usepackage{amssymb}
\usepackage{graphicx}
\usepackage{float}
\usepackage{subfigure}
\usepackage{dcolumn}
\usepackage{bm}
\usepackage{bbm}
\usepackage{amsthm}
\usepackage{thmtools}

\usepackage{mathtools}
\usepackage{physics}
\usepackage{binarytree}
\usepackage{tikz,pgfplots}
\usepackage{verbatim}
\usepackage{pgfplots}
\usepackage{placeins}
\usepackage{algorithm2e}
\usetikzlibrary{decorations.pathreplacing}
\usetikzlibrary{arrows.meta}
\usetikzlibrary{graphs}
\usepackage{xcolor}
\usepackage{rotating}
\usepackage{yquant}
\usepackage{tabularray}
\usepackage{tabularx}
\usepackage{times}
\usepackage[colorlinks,bookmarks=true,citecolor=green,linkcolor=blue,urlcolor=blue]{hyperref}
\RestyleAlgo{ruled}
\pgfplotsset{compat=1.16}

\begin{document}

\title{Variational preparation and characterization of chiral spin liquids in quantum circuits}

\author{Zi-Yang Zhang}
\affiliation{Center for Neutron Science and Technology, Guangdong Provincial Key Laboratory of Magnetoelectric Physics and Devices, School of Physics, Sun Yat-sen University, Guangzhou 510275, China}

\author{Donghoon Kim}
\affiliation{Analytical Quantum Complexity RIKEN Hakubi Research Team, RIKEN Center for Quantum Computing (RQC), Wako, Saitama 351-0198, Japan}

\author{Ji-Yao Chen}
\email{chenjiy3@mail.sysu.edu.cn}
\affiliation{Center for Neutron Science and Technology, Guangdong Provincial Key Laboratory of Magnetoelectric Physics and Devices, School of Physics, Sun Yat-sen University, Guangzhou 510275, China}

\date{\today}

\begin{abstract}

Quantum circuits have been shown to be a fertile ground for realizing long-range entangled phases of matter. While various quantum double models with non-chiral topological order have been theoretically investigated and experimentally implemented, the realization and characterization of chiral topological phases have remained less explored. Here we show that chiral topological phases in spin systems, i.e., chiral spin liquids, can be prepared in quantum circuits using the variational quantum eigensolver (VQE) framework. On top of the VQE ground state, signatures of the chiral topological order are revealed 
using the recently proposed tangent space excitation ansatz for quantum circuits. We show that, both topological ground state degeneracy and the chiral edge mode can be faithfully captured by this approach.
We demonstrate our approach using the Kitaev honeycomb model, finding excellent agreement of low-energy excitation spectrum on quantum circuits with exact solution in all topological sectors. Further applying this approach to a non-exactly solvable chiral spin liquid model on square lattice, the results suggest this approach works well even when the topological sectors are not exactly known.

\end{abstract}

\maketitle

\section{Introduction}


With the rapid advancement of quantum hardware, exotic many-body phases of matter are progressively becoming experimentally accessible~\cite{Satzinger2021,Semeghini2021,Ebadi2021,Hoke2023,Cuadra2025,Cochran2025,Liu2025,Alam2025}. Among them, topological phases with long-range entanglement are particularly intriguing, for which digital quantum devices provide a versatile platform to create, braid and fuse their anyonic excitations~\cite{Andersen2023,Iqbal2024,Xu2024}. Focusing on two spatial dimension, topologically ordered phases can be roughly divided into two classes -- non-chiral and chiral cases~\cite{Wen2017}. For non-chiral topologically ordered phases, especially at the renormalization fixed point with zero correlation length, the state preparation with sequential unitary circuits has been relatively well understood, with the notable example of toric code being realized on various quantum devices~\cite{Satzinger2021,Liu2022,Wei2022}. Further augmenting the unitary circuits with measurement and feedforward, constant depth circuit has been proposed and implemented to prepare abelian and non-abelian non-chiral topological states~\cite{Piroli2021,Tantivasadakarn2024,Iqbal2024,Zhang2024}. In contrast, chiral topological states are much less explored on quantum devices, except for recent experimental progress on the paradigmatic Kitaev honeycomb model~\cite{Kitaev2006,Will2025,Evered2025}.

Chiral topological phases support not only bulk anyonic excitations, but also edge modes on the boundary that propagate in one direction~\cite{Kalmeyer1987,Wen1989,Wen1990,Wen1991}. Contrary to non-chiral cases, chiral topological phase does not have a renormalization fixed point with zero correlation length. Thus the state preparation cannot be achieved solely via measurement and feedforward. Notable proposals for realizing chiral phases include circuits based on multiscale entanglement renormalization group (MERA)~\cite{Chu2023}, sequential generation circuits~\cite{Chen2025b}, and in a non-equilibrium way using Floquet engineering~\cite{Kalinowski2023,Mambrini2024}. While both MERA circuits and sequential generation require analog simulation with short-range or longer-range Hamiltonian as part of the scheme and thus unfeasible for current devices, the Floquet engineering scheme appears to be more accessible to current quantum devices. Indeed, experimental works on chiral topological phase in Kitaev honeycomb model~\cite{Evered2025,Will2025} rely on a Floquet unitary circuit. However, it is not entirely clear the full power of this type of circuit, whether the circuit is generalizable to generic chiral topological phases, and what else one could learn about chiral phases using this type of circuits.

A more stringent issue on chiral topological phases is the probe on quantum devices. As stabilizer formalism does not fully capture the chiral topological order, state characterization on quantum devices is more involved, for which recent experiments on Kitaev honeycomb model have used the chiral edge mode to reveal the chiral nature~\cite{Will2025}, or through learning Chern number of the associated fermion band~\cite{Evered2025}. Both two approaches are limited for providing a decisive evidence for generic chiral topological phases. As topological phases are typically characterized by their entanglement properties, an efficient way of classically characterizing chiral phase is through entanglement spectrum (ES)~\cite{Li2008}, which is available in tensor networks~\cite{Cincio2013,Cirac2011}. However, on quantum devices, measuring ES through state tomography requires an exponential overhead and is challenging~\cite{Kokail2021}. Another useful way of identifying chiral topological order is through measuring momentum polarization~\cite{Tu2013}, which provides partial information of the order. Thus, a different way of efficiently characterizing the chiral state on quantum devices is needed.

From an experimental point of view, many-body phases of matter are traditionally probed by their responses to external fields, e.g., dynamical structure factor via neutron scattering experiments~\cite{Norman2016,Banerjee2017}. The relevant physical observables are then determined by the excited state of a given system. On quantum devices, although several works have studied ground state preparation of topological spin liquids and their anyon excitations~\cite{Iqbal2024,Xu2024}, with efforts on developing quantum algorithms for excited states~\cite{Higgott2019,Nakanishi2019,Farrell2025,Velury2025}, much less is known for the low-lying spectrum of systems with topological order. Interestingly, topologically ordered phases, by definition, have manifold dependent ground state degeneracy, and thus exhibit different spectrum with changing boundary conditions. Thus it appears tempting to both characterize and reveal experimental signatures of chiral topological phase through its low-energy spectrum on quantum devices.

In this work, we focus on the chiral topological phases in spin systems, namely the chiral spin liquids (CSLs)~\cite{Kalmeyer1987,Wen1989,Savary2017}. Turning the Floquet unitary circuit into a variational quantum eigensolver (VQE)~\cite{McArdle2020,Cerezo2021,Tilly2022}, we show through examples that, generic chiral topological states can be prepared. Moreover, for CSL in Kitaev honeycomb model, we show all the degenerate ground states can be exactly prepared (in machine precision) with a small circuit depth. As Kitaev honeycomb model is equivalent to a free fermion model coupled to static $\mathbb{Z}_2$ gauge field, it is suggestive that all representative models in Kitaev's $16$-fold way~\cite{Kitaev2006,Chulliparambil2020} can be prepared efficiently in this way. Further using the recently developed tangent space excitation ansatz for quantum circuits~\cite{Chen2025a}, evidence for chiral topological order can be found through computing low-energy excitation spectrum of the given system on both torus and disk geometries. Notably, the computational cost of this approach scales with a low-order polynomial with the circuit depth and system size, making it attractive for near term quantum devices. Different from the previous works of simulating excitations with projected entangled-pair state (PEPS)~\cite{Vanderstraeten2019a}, where the calculation is performed on the infinite plane, here we tackle the finite system and each individual topological sector can be clearly identified.

This manuscript is organized as follows. We start with the highly accurate ground states for Kitaev honeycomb model through VQE in Sec.~\ref{sec:VQE_Kitaev}, and present the excitation spectrum of this model in both zero vortex and non-zero vortex sectors in Sec.~\ref{sec:Kitaev_excitation}. These two sections serve as a useful benchmark for our approach. Then in Sec.~\ref{sec:Square_CSL_torus} we use this approach to study a $\mathrm{SU}(2)_1$ CSL in a chiral Heisenberg antiferromagnet on the square lattice~\cite{Nielsen2013}, and identify topological ground state degeneracy on torus geometry through the low-energy spectrum. We further investigate the excitation spectrum of the model on open disk geometry in Sec.~\ref{sec:Square_CSL_disk}, finding smoking gun evidence for the gapless edge modes described by a $\mathrm{SU}(2)_1$ conformal field theory (CFT)~\cite{Francesco1997}. Different from the Kitaev CSL, the associated 1-form symmetry in the $\mathrm{SU}(2)_1$ CSL case is emergent instead of being exact, which thus demonstrates the usefulness of our approach. We conclude with a discussion of other possible circuit layout in Sec.~\ref{sec:Outlook}. The appendix provides the analytical solution of Kitaev honeycomb model and several necessary details on quantum circuits.

\section{VQE for Kitaev honeycomb model}
\label{sec:VQE_Kitaev}

\subsection{Kitaev honeycomb model}
\label{subsec:Kitaev_model}

\begin{figure}[hptb]
\centering
\includegraphics[width=0.95\columnwidth]{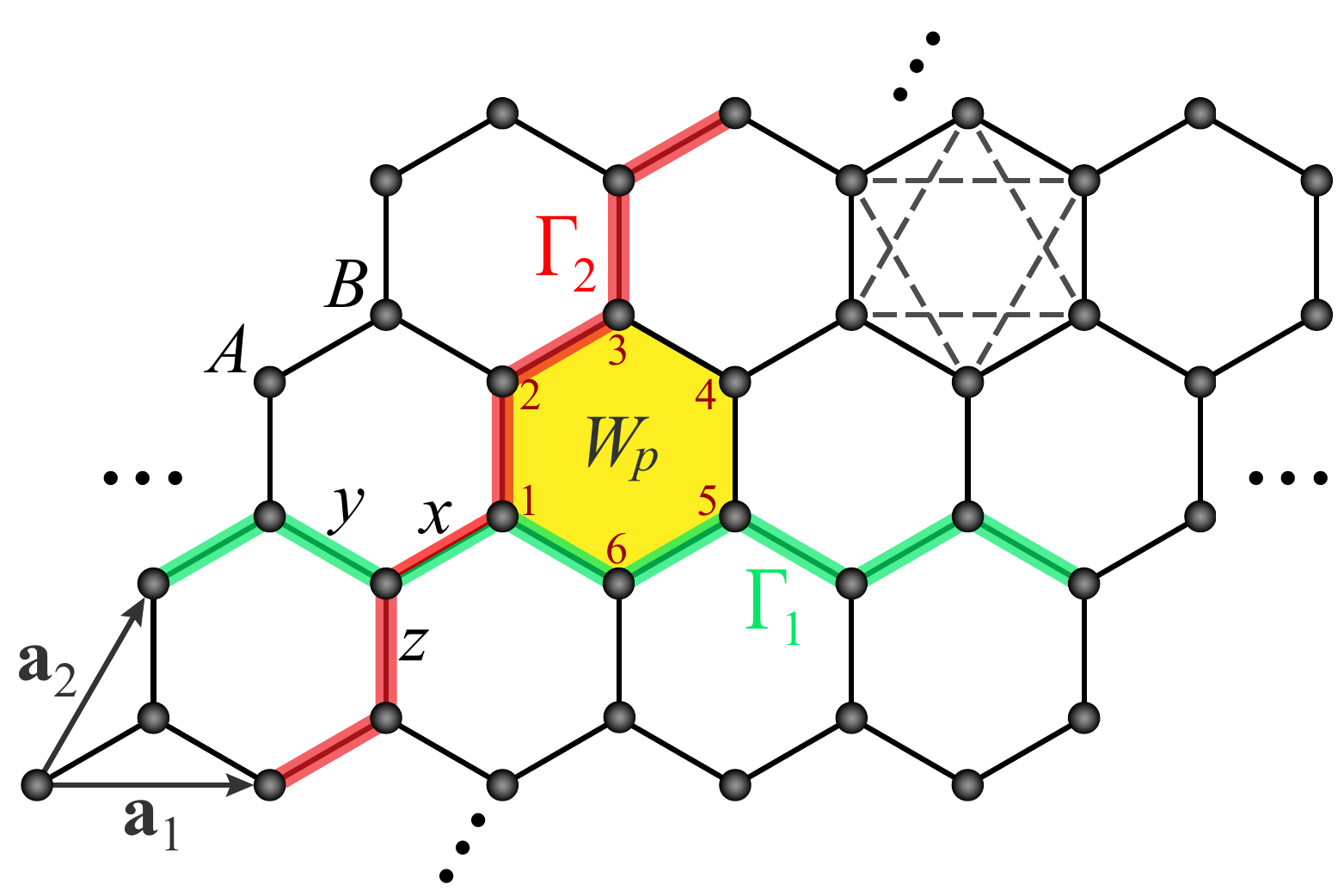}
\caption{Schematic for Kitaev honeycomb model. The lattice is generated by translating a two-site unit cell along $\bf{a}_1$ and $\bf{a}_2$ directions, and contains $A,B$ sublattices. Depending on the orientation, links of the lattice are labeled as $x,y,z$. Six small triangles in each hexagon are illustrated in upper right. The site label for plaquette operator $W_p$ is shown in the yellow hexagon. The non-contractible loop $\Gamma_1$ ($\Gamma_2$) along $\bf{a}_1$ ($\bf{a}_2$) direction is shown in green (red).
}
\label{fig:Kitaev_model}
\end{figure}

We start with the exactly solvable Kitaev honeycomb model~\cite{Kitaev2006}, which provides a suitable benchmark for our approach. 
On each site of the honeycomb lattice, we place a spin-$1/2$ degree of freedom. 
Depending on the orientation, we label the links of the lattice by $\alpha=x,y,z$, as shown in Fig.~\ref{fig:Kitaev_model}. The Hamiltonian is given by:
\begin{equation}
    H_{\mathrm{Kitaev}} = \sum_{\alpha}J_{\alpha}\sum_{\langle i,j\rangle\in\alpha-{\mathrm{links}}}\sigma_{i}^{\alpha}\sigma_{j}^{\alpha} + K \sum_{\langle ijk\rangle\in \triangle}\sigma_i^x\sigma_j^y\sigma_k^z,
\label{eq:Kitaev_model}
\end{equation}
where $\sigma^{\alpha} (\alpha=x,y,z)$ are Pauli matrices. The $J_{\alpha}$ terms are nearest-neighbor Ising type interactions on each link, whose form depends on the link type. The $K$ term acts on all small triangles of each hexagon, each of which contains two neighboring links connected by one site. The local spin operator on the shared site is determined by the link type absent in the corresponding triangle, while the other two follow the type of the links they act on. To illustrate, $K$ terms in one hexagon are shown in Fig.~\ref{fig:Kitaev_model}.

The Kitaev honeycomb model has a rich phase diagram. With $K=0$ in Eq.~\eqref{eq:Kitaev_model}, the model realizes both gapped and gapless non-chiral spin liquids in the phase diagram~\cite{Kitaev2006}. The latter is protected by the time-reversal symmetry. As pointed out by Kitaev in Ref.~\cite{Kitaev2006},
a nonzero $K$ term can break the time-reversal symmetry, while preserving the exact solvability via Majorana fermion representation. In the following we will focus on the Hamiltonian parameter $J_x=J_y=J_z=-1$, $K=0.2$, for which the model is known to realize a chiral spin liquid, supporting the Ising topological order~\cite{Kitaev2006,Lahtinen2008}.

The model has a set of locally conserved quantities, which are akin to the $\mathbb{Z}_2$ lattice gauge theory and useful for understanding its properties and analyzing the spectrum. Denoting the linear size along $\bf{a}_1$ ($\bf{a}_2$) direction (shown in Fig.~\ref{fig:Kitaev_model}) as $L_1$ ($L_2$), with periodic boundary condition imposed in both directions, the number of hexagons (sites) is $N=L_1L_2$ ($N_s=2L_1L_2$). 
On hexagon $p$ ($p=1,2,\ldots,N$), one can define a plaquette operator $W_p=\sigma_1^x\sigma_2^y\sigma_3^z\sigma_4^x\sigma_5^y\sigma_6^z$ (see Fig.~\ref{fig:Kitaev_model} for the site label), which commutes with the Hamiltonian and also mutually commutes. Thus one can use the eigenvalues of all plaquette operators to label eigenstates of the model.
The plaquette operator satisfies $W_p^2=1$ with eigenvalues being $\pm 1$. Based on the exact solution, it is known that the ground state lives in the sector where all $W_p$'s take eigenvalue $+1$, which, following the convention~\cite{Kitaev2006,Lahtinen2008}, is called the flux free sector. If on a plaquette $p$ we have $\langle W_p\rangle=-1$, then we call there is a vortex on this plaquette. On torus geometry, the product of all $W_p$'s equals to 1, meaning that vortex appears in pairs.

In addition to the local conserved quantities, one can further define Wilson loop operators $\Phi_1$ and $\Phi_2$ on torus, taking the form: $\Phi_1 = \otimes_{i\in \Gamma_1}\sigma_i^z$, $\Phi_2=\otimes_{i\in\Gamma_2}\sigma_i^y$. Here $\Gamma_1$ ($\Gamma_2$) is the non-contractible loop along $\bf{a}_1$ ($\bf{a}_2$) direction, graphically shown in Fig.~\ref{fig:Kitaev_model}. Both $\Phi_1$, $\Phi_2$ commute with the Hamiltonian and also mutually commute, satisfying $\Phi_1^2=\Phi_2^2=1$. Moreover, the Wilson loop operators also commute with all the $W_p$'s, with the inter relation 
\begin{equation}
\Phi_{1,y}\Phi_{1,y+1}=\prod_{p} W_{p},\quad \Phi_{2,x}\Phi_{2,x+1}=\prod_{p} W_{p}.
\label{eq:1_form_symmetry}
\end{equation}
Here we have used $x$ ($y$) to denote the coordinate along $\bf{a}_1$ ($\bf{a}_2$) direction, and both products run over all hexagons between the two non-contractible loops. In a modern terminology, the set of all $\Phi_1, \Phi_2, W_p$ forms a so-called 1-form symmetry~\cite{Ellison2023}, which guarantees deconfined fermionic excitations in a gapped spin-$1/2$ system~\cite{Liu2024} (see also related discussions in Ref.~\cite{Chen2024b}). For notational clarity, in the following we use $\phi_1$ ($\phi_2$) to denote the eigenvalue of $\Phi_1$ ($\Phi_2$), respectively.

It is useful to introduce another set of loop operators for the Kitaev honeycomb model here. In alignment with the $\mathbb{Z}_2$ toric code where one has anti-commuting loop operators acting on the logical space, here we define a set of loop operators which can flip the eigenvalues of Wilson loop operators while preserving the vortex configuration. Specifically, we consider two loop operators $U_{a} (a=1,2)$ along the non-contractible loops on torus, with the following algebraic relations:
\begin{equation}
\left[U_{a},W_{p}\right] = 0,\quad \left[U_{a},\Phi_{a}\right]=0,\quad \left\{U_{a},\Phi_{b}\right\} = 0,
\label{eq:logical_X_operator}
\end{equation}
where $a,b=1,2$ and $a\neq b$. While the forms of $U_a (a=1,2)$ are not unique, a convenient choice is given by 
\begin{equation}
U_1=\prod_{i\in\Gamma_1\cap A}\sigma_i^z,\quad
U_2=\prod_{i\in\Gamma_2\cap A}\sigma_i^y,
\label{eq:loop_X_expression}
\end{equation}
where $A$ represents the $A$-sublattice of the honeycomb lattice. It is easy to see the support of $U_1$ ($U_2$) operator is actually half of the sites on non-contractible loop $\Gamma_1$ ($\Gamma_2$). Note that, the $U_1$ and $U_2$ constructed in this way anti-commute: $\{U_1,U_2\}=0$. This may appear odd for the toric code phase of this model. However, one can simply redefine $U_2$ as $\tilde{U}_2=U_2\Phi_2$ to make it commute with $U_1$. In the following we keep Eq.~\eqref{eq:loop_X_expression} as it is without causing any inconvenience.

On torus, the CSL phase realized in the Kitaev honeycomb model is known to have three topologically degenerate ground states. From fermion solution (presented in Appendix~\ref{app:Majorana}) and exact diagonalization (ED), one can find that the three degenerate ground states all have $\langle W_p\rangle=1$ and zero momentum. Further, the three ground states have distinct sets of eigenvalues for Wilson loop operators: the absolute ground state on finite torus has $(\phi_1,\phi_2)=(1,1)$, while the other two topological degenerate ground states each has $(\phi_1,\phi_2)=(1,-1)$ and $(\phi_1,\phi_2)=(-1,1)$, respectively. Notice that the lowest energy state in $(\phi_1,\phi_2)=(-1,-1)$ sector is an excited state with nonzero momentum, the reason of which is explained in Appendix~\ref{app:Majorana}.
With these conserved quantities identified, we are now ready to prepare the ground state on quantum circuits.

\subsection{Hamiltonian variational ansatz for Kitaev CSL}
\label{subsec:HVA_Kitaev}

Guided by the conserved quantities, we can now prepare the variational ground state on quantum circuits, using suitable initial states that are common eigenstates of $W_p,\Phi_1,\Phi_2$ with specific eigenvalues. We further require the initial states to be translationally invariant with momentum zero, in line with the quantum numbers of exact ground states. Since $W_p,\Phi_1,\Phi_2$ can be viewed as stabilizers, a simple way to prepare such initial states would be through projective measurement on a product state and feedforward, which has been realized experimentally~\cite{Evered2025}. In Appendix~\ref{app:ini_state}, we provide an alternative approach for initial state preparation using a sequential circuit. This will be useful for probing excitation spectrum of this model with high accuracy. In the following we assume the initial states are available and fixed, and present the protocol for variational ground states.

Although there have been illustrative analytical studies on preparing the chiral topological states on quantum circuits~\cite{Chu2023,Chen2025b}, and on preparing ground state of Kitaev honeycomb model using its free fermion nature~\cite{Schmoll2017,Jahin2022,Park2025}, the circuit architecture are intricate and not generalizable for other chiral phases. Here we resort to the VQE framework, taking further insight from early studies of Kitaev CSL using Floquet engineering~\cite{Kalinowski2023,Sun2023}. The circuit ansatz we used for searching ground states is a variant of the Hamiltonian Variational Ansatz (HVA)~\cite{Wecker2015,Ho2019,Wiersema2020}, with restriction of keeping only the two-body terms in the Hamiltonian. This is reasonable since one can find that the three-spin terms in the model Eq.~\eqref{eq:Kitaev_model} can be obtained from commutators of the two-site terms~\cite{Kalinowski2023}. Indeed, Eq.~\eqref{eq:Kitaev_model} can be rewritten in the following form:
\begin{equation}
\begin{split}
    H_{\mathrm{Kitaev}} = \sum_{\alpha=x,y,z}J_{\alpha}H^{\alpha} & -\frac{iK}{2}\big( [H^x,H^y]+ \\
    &[H^y,H^z]+[H^z,H^x] \big),
\end{split}
\label{eq:Kitaev_commutator}
\end{equation}
where we have denoted $H^{\alpha}=\sum_{\langle ij \rangle\in\alpha-{\mathrm{links}}}\sigma_i^{\alpha}\sigma_j^{\alpha}$. Based on Eq.~\eqref{eq:Kitaev_commutator}, it was found the time evolution with $H_{\mathrm{Kitaev}}$ can be approximated using only the two-body terms in the Floquet engineering framework~\cite{Kalinowski2023}.

\begin{figure}[hptb]
\centering
\includegraphics[width=0.95\columnwidth]{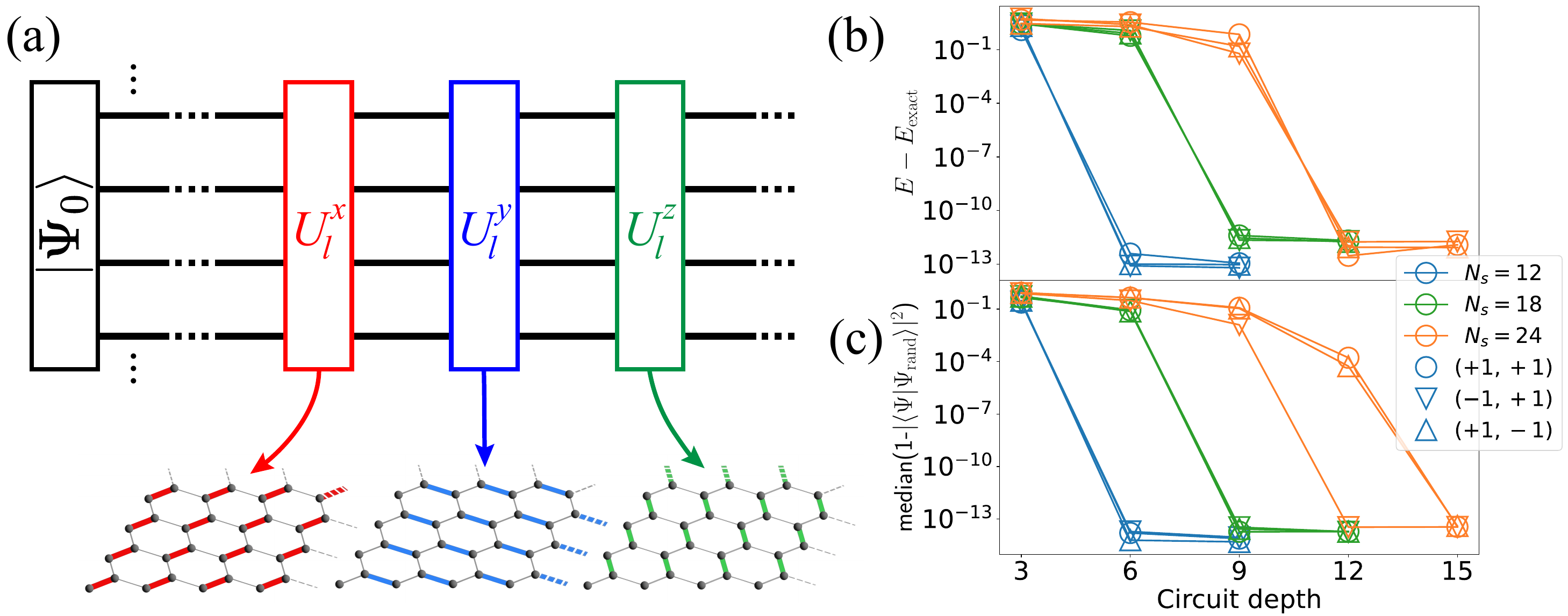}
\caption{(a) Schematics of HVA for Kitaev honeycomb model. For a $(\phi_1,\phi_2)$ sector, $|\Psi_{0}\rangle$ is a proper initial state with well-defined conserved quantities. The links where $U_{l}^{\alpha}$'s act are shown in the lower panel, highlighted with the corresponding color. In (b) and (c) we show the variational power of HVA for system sizes $(L_1,L_2)=(3,2),(3,3),(4,3)$ with number of sites $N_s=2L_1L_2$. (b) For all three $(\phi_1,\phi_2)$ sectors, comparing with exact ground state, the energy error of optimized variational state decreases with circuit depth, reaching a value close to zero. (c) For these system sizes, the state generated by a circuit with random parameters $\{\bm{\theta}\}$ and depth $D=12,15,18$ can be approached with high precision using optimized circuits with lower depth $D-3$, where a sudden drop in median of infidelity over 100 samples is observed.}
\label{fig:Kitaev_HVA}
\end{figure}

The circuit ansatz reads:
\begin{equation}
    U({\bm{\theta}}) = \prod_{l=1}^{D/3}\mathrm{e}^{-i\theta_{l}^{z}H^z} \mathrm{e}^{-i\theta_{l}^{y}H^y} \mathrm{e}^{-i\theta_{l}^{x}H^x},
\label{eq:Kitaev_HVA}
\end{equation}
where $\bm{\theta}=\{\theta_l^{\alpha} (l=1,2,\ldots,D/3,\alpha=x,y,z)\}$ are variational parameters of the ansatz with circuit depth $D$ ($D$ is a multiple of 3). Here each layer only acts on one type of nearest neighbor links, with the gate generated by time evolution of corresponding two-body terms in Eq.~\eqref{eq:Kitaev_model}. We have further taken the parameters of all gates in one layer to be the same, so that translation symmetry is preserved by the circuit. The variational state in the zero vortex sector with $\Phi_1,\Phi_2$ eigenvalues being $(\phi_1,\phi_2)$ is then given by 
\begin{equation}
    |\Psi\rangle=U(\bm{\theta})|\Psi_0\rangle,
\label{eq:Kitaev_ground_state}
\end{equation}
where $|\Psi_0\rangle$ is the initial state of the corresponding sector, prepared either by measurement and feedforward, or by the sequential circuit in Appendix~\ref{app:ini_state}. A pictorial representation of this circuit is shown in Fig.~\ref{fig:Kitaev_HVA}(a). It is easy to see that the conserved quantities in $|\Psi_0\rangle$ are preserved by the HVA circuit.

Using this scheme, we have optimized the HVA for three system sizes $(L_1,L_2)=(3,2),(3,3),(4,3)$, respectively. For all three $(\phi_1,\phi_2)$ sectors containing the topologically degenerate ground states, comparing with the exact results, the energy error of the optimized variational state decreases with circuit depth. Interestingly, as shown in Fig.~\ref{fig:Kitaev_HVA}(b), the errors reach a value close to zero with a small depth, beyond which, the errors remain essentially unchanged due to finite precision in the optimizer. For the system sizes we have studied, the threshold circuit depth increases with system size, although based on the results available, it is not clear whether this value increases linearly with system sizes. Similar behavior in energy error was also observed in HVA for 1D transverse field Ising model (TFIM)~\cite{Ho2019} and the Kitaev honeycomb model without chiral term~\cite{Bespalova2021}.

To obtain more understanding of the variational power of the HVA circuit, we consider states generated by Eq.~\eqref{eq:Kitaev_ground_state} with parameters $\{\bm{\theta}\}$ chosen uniformly random from $[-\pi,\pi]$. We further set the depth $D=12,15,18$ for system size $(L_1,L_2)=(3,2),(3,3),(4,3)$, respectively, which are sufficient for optimized states to approach the exact ground states (see Fig.~\ref{fig:Kitaev_HVA}(b)). Now we optimize circuits with smaller depth to approximate the random states, using infidelity as the cost function. As shown in Fig.~\ref{fig:Kitaev_HVA}(c), the median of infidelity over 100 realizations for all system sizes with three sectors decreases to a value close to zero with circuit depth $D-3$, consistent with results shown in Fig.~\ref{fig:Kitaev_HVA}(b). The nonzero error for $N_s=24$ site at depth $D=12$ in the random case suggests this depth is not sufficient for random states, although the ground state can be approximately found with high accuracy using this circuit depth. The numerical results for random states imply that any state generated by Eq.~\eqref{eq:Kitaev_HVA} with arbitrary circuit depth can be recursively reduced to a circuit with depth $D=9,12,15$ for $N_s=12,18,24$, respectively. A closely related behavior with HVA for 1D TFIM was observed in Ref.~\cite{Ho2019}. As both TFIM and Kitaev honeycomb model can be mapped to free fermion models, we suspect that this behavior is generic for free fermions, and leave a detailed analytical study to future works.

Before moving further, it is interesting to make a comparison of the circuit architecture we use with sequential adiabatic generation introduced in Ref.~\cite{Chen2025b} and circuits based on  MERA in Ref.~\cite{Chu2023}. Recall that using the Majorana fermion representation, the Kitaev honeycomb model is equivalent to a topological superconductor coupled to a static $\mathbb{Z}_2$ gauge field~\cite{Kitaev2006}. All three approaches use a sequential circuit for the $\mathbb{Z}_2$ gauge part, and the difference lies in how to realize the free fermion part. While Ref.~\cite{Chu2023} has used a quasi-local evolution to realize the chiral state, Ref.~\cite{Chen2025b} proposed a sequential adiabatic evolution to turn trivial atomic insulator into a topological insulator. Thus the gates in these two approaches are not strictly local, in contrast to our approach.

One can also understand the difference between our approach and sequential circuits from the tensor network point of view, where sequential circuit is obtained by replacing the Hamiltonian adiabatic evolution in the sequential generation~\cite{Chen2025b} with strictly local gates. The HVA circuit can be naturally represented as a PEPS~\cite{Cirac2021,Xiang2023} (although the bond dimension may increase with system size), while sequential circuit leads to an isometric PEPS. As PEPS has been demonstrated to represent chiral states faithfully in several works~\cite{Poilblanc2017b,Chen2018,Chen2020,Hasik2022,Niu2022}, to our knowledge, related results using isometric PEPS have been lacking. Instead, a recent work has established no-go theorems for chiral phase in isometric PEPS~\cite{Fan2025}, suggesting that isometric PEPS, or equivalently sequential circuit, has certain limitations in describing chiral phases.

\section{Tangent space excitation of Kitaev CSL}
\label{sec:Kitaev_excitation}

With the variational ground states available, we are now ready to explore the properties of the phase and characterize the low-energy excitation spectrum. This provides not only a way to study the anyon physics contained in this model, but also a testbed for our approach. Note that, although the excitation spectrum of Kitaev honeycomb model (with potentially additional terms) have been investigated using PEPS based excitation ansatz~\cite{Tan2024,Wang2025}, the results are all on infinite plane, without considering topological sectors explicitly. In contrast, our approach with quantum circuits will focus on finite size system and further take into account the effects of Wilson loops.

From the exact solution, it is known that the CSL phase of this model contains three types of anyonic excitations, including the trivial particle $I$, the fermion excitation $\psi$, and the Ising anyon $\sigma$ which is a $\mathbb{Z}_2$ vortex with a Majorana fermion trapped in it~\cite{Kitaev2006}. The Ising anyons satisfy the fusion rule $\sigma\times\sigma=I+\psi$, realizing a non-abelian topological order. Generically, the anyon excitations come in pairs, forming a continuous spectrum in the thermodynamic limit, although special features exist in the Kitaev honeycomb model which we will discuss later. Due to conservation of $W_p$, the vortices are localized and static. Thus the excitation spectrum can be labeled by the vortex configuration.

In the vortex free sector, the excitations contain fermions with a characteristic fermion gap~\cite{Lahtinen2008}. With the presence of translation invariance, we can further label the excited states using momentum quantum numbers and eigenvalues of the Wilson loop operators. In the sector with vortices, the characteristic quantity is the vortex gap, which is the energy difference between lowest energy eigenstate with vortex and the ground state~\cite{Lahtinen2008}. Since different vortex configurations are mutually orthogonal, it is more convenient to work with a fixed vortex configuration. In that case, the translation invariance is broken, and the fermions can hop and form pairs in the background of the $\mathbb{Z}_2$ gauge field.

To access the low energy excitations of this system, we resort to the tangent space excitation ansatz for quantum circuits, which, based on locality of the interactions, is a native approach for capturing quasi-particle properties on quantum computers~\cite{Chen2025a}. In this approach, one constructs excited states by inserting local operators into the ground state circuit, the collection of which forms the tangent space of the corresponding tensor network state manifold at the variational optimum and is a faithful subspace for low-energy excitations. Here we will not elaborate on the difference between this approach and PEPS based excitation ansatz, and refers to the original work~\cite{Chen2025a} for further discussions. In principle, the excitation ansatz is designed to capture the single-particle excitation (in the spin language). Thus it is not known {\it a priori} how well the performance of the ansatz will be for the CSL we are studying.

\begin{table}[htb]
\renewcommand\arraystretch{1.5}
\caption{Loop operators transforming state from initial $(\phi_1,\phi_2)$ sector in the top row into target $(\phi'_1,\phi'_2)$ sector in the other rows.}
\label{table:loop_operator}
\begin{tabularx}{0.48\textwidth}{l l l l}
	\hline
	\hline
	$(\phi_1,\phi_2)$ \qquad& $(+1,+1)$ \qquad\qquad & $(+1,-1)$ \qquad\qquad & $(-1,+1)$ \qquad\qquad \\
	\hline
	$(+1,+1)$ \qquad\qquad & \qquad $I$ \qquad & \qquad $U_1$ \qquad & \qquad $U_2$ \qquad \\
    $(+1,-1)$ \qquad\qquad & \qquad $U_1$ \qquad & \qquad $I$ \qquad & \qquad $U_1U_2$ \qquad \\
    $(-1,+1)$ \qquad\qquad & \qquad $U_2$ \qquad & \qquad $U_1U_2$ \qquad & \qquad $I$ \qquad \\
    $(-1,-1)$ \qquad\qquad & \qquad $U_1U_2$ \qquad & \qquad $U_2$ \qquad & \qquad $U_1$ \qquad \vspace{1mm}\\
	\hline
	\hline
\end{tabularx}
\end{table}

In addition, since the ground state has three-fold topological degeneracy, we will consider all three ground states, together with the loop operators $U_{a} (a=1,2)$ introduced in Eq.~\eqref{eq:loop_X_expression} to construct the subspace for excitations. Table~\ref{table:loop_operator} shows how the ground state in one $(\phi_1,\phi_2)$ sector is transformed into target $(\phi'_1,\phi'_2)$ sector. We found this approach can significantly improve the accuracy for certain sectors. 
For the case using only one ground state for the corresponding sector, see Appendix~\ref{app:Numerical_details}. 
Since the excitation ansatz for vortex free sectors and sectors with vortex have slightly different form due to the translation symmetry, we now present them separately.

\subsection{Vortex free sector}
\label{subsec:zero_vortex}

\begin{figure*}[hptb]
    \centering
    \includegraphics[width=1.9\columnwidth]{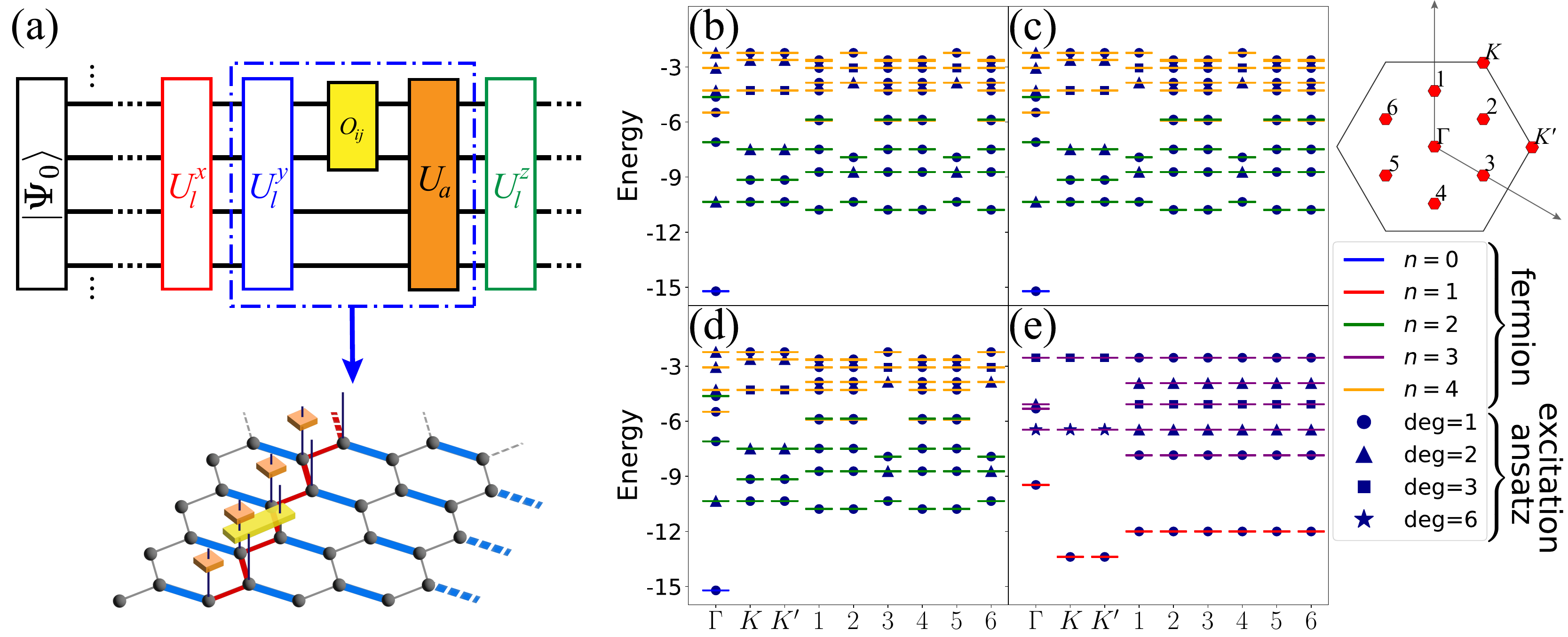}
    \caption{(a) Schematics of one basis state for excitations on quantum circuits, with the detailed implementation illustrated below. (b),(c),(d) and (e) show the low-energy spectrum of a $(L_1,L_2)=(3,3)$ system in zero flux sectors, with $(\phi_1,\phi_2)=(1,1), (-1,1), (1,-1), (-1,-1)$, respectively. The available momenta are indicated by red filled dots in the first Brillouin zone shown on the right. Lines show the spectrum obtained with exact fermion solution, which is labeled by the number of occupied fermion modes $n$ on top of the fermion ground state. Filled symbols in (b)-(e) represent spectrum from our variational ansatz with three ground states, where distinct symbols represent energy levels with different degeneracy. In all sectors, the variational results, including the level degeneracy, agree with exact results over a wide range of energy.
    }
\label{fig:Kitaev_zero_vortex}
\end{figure*}

Using the relations between Wilson loop operators and plaquette operator $W_p$ (Eq.~\eqref{eq:1_form_symmetry}), the Hamiltonian in the vortex free sector can be diagonalized simultaneously with the translation operators and Wilson loop operators, giving rise to momentum $\bm{K}$ and $\phi_1$, $\phi_2$ as good quantum numbers. From the above VQE calculation, we have obtained three topologically degenerate ground states, lying in sectors $(1,1)$, $(1,-1)$, and $(-1,1)$, respectively, with the sector $(-1, -1)$ untouched.

Although obtaining the lowest energy state with nonzero momentum in $(-1,-1)$ sector is possible by supplying HVA with a suitably entangled initial state, we do not consider it in this work. Instead, to reveal the fact that this state is an excited state of the model, we will approach this sector using the tangent space excitation ansatz. To reach the sector $(-1,-1)$, we use the two loop operators $U_{a} (a=1,2)$ to flip the eigenvalues of corresponding Wilson loop operators, with vortex configuration unchanged. In the fermion language, the $U_a$ operators change the fermion boundary conditions, while also inserting fermions to ground state due to non-commutation between $U_a$ and the Hamiltonian. Note that, although we have fixed the form of $U_{a}$ as in Eq.~\eqref{eq:loop_X_expression}, their locations are not unique and can be inserted at arbitrary depth of the circuit, giving some freedom in the basis construction for excited states.

For the Kitaev honeycomb model, the conserved quantities impose strong constraints to the inserted local operators. In vortex free sectors, the local operators need to commute with the $W_p$'s and $\Phi_1, \Phi_2$. Since single Pauli operator does not commute with all the $W_p$'s surrounded, and the dimension of the excitation subspace grows exponentially with the size of the local operator support, in the following we consider two-site operators $O_{ij}$ acting on nearest-neighbor sites $i,j$ to create excitations. Physically, this can be interpreted as changing the fermion configuration with fixed $\mathbb{Z}_2$ gauge configuration. Specifically, at a given depth, on each $\alpha$-link ($\alpha=x,y,z$), the local operator we inserted is the local Hamiltonian term in Eq.~\eqref{eq:Kitaev_model}, i.e., $O_{ij}=\sigma_i^{\alpha}\sigma_j^{\alpha}$ for $\alpha$-link connecting site $i$ and $j$. With this setup, we now describe the basis in the tangent space excitation ansatz~\cite{Chen2025a}.

Taking one basis as an example, we insert the operator $O_{ij}$ at arbitrary depth of the ground state circuit (see Fig.~\ref{fig:Kitaev_zero_vortex}(a) for illustration), which is possibly followed by $U_{a}(a=1,2)$ loop operator to enforce the desired $(\phi_1,\phi_2)$ sector. To achieve a high accuracy, we find that it is best to choose the loop close to the link $O_{ij}$ acting on. Specifically, for an $x$ or $y (z)$ link operator, the $\Gamma_1(\Gamma_2)$ loop contains this link, while for a $z (y)$ link operator, $B$ site of the link is contained in the corresponding loop. When both $U_1$ and $U_2$ are needed, $U_2$ is applied first, followed by $U_1$. Moreover, as the ground state circuit are translationally invariant, we take a momentum superposition of the circuit configuration, so that the momentum quantum number is manifest. Thus one possible basis is given by the following form:
\begin{equation}
\begin{split}
|\psi({\bm{K}})\rangle=\sum_{n=0}^{L_2-1}\sum_{m=0}^{L_1-1}&\mathrm{e}^{-i(k_1m+k_2n)}T_2^nT_1^m\\
&U(\bm{\theta_2})U_1OU(\bm{\theta_1})|\Psi_0\rangle,
\end{split}
\label{eq:Kitaev_zero_vortex}
\end{equation}
where $|\Phi_0\rangle$ is one of the initial states with good quantum numbers, $O$ is the perturbation operator inserted on one link and $U(\bm{\theta_1})$ ($U(\bm{\theta_2})$) represents part of the ground state circuits before (after) the perturbation layer. The normalization of this state is omitted and left to the optimization process. The eigenvalue of translation operator $T_1$ is then given by $\mathrm{e}^{i2\pi k_1/L_1}$ $(k_1=0,\ldots,L_1-1)$. Similar expression holds for $T_2$.

After obtaining the basis states $\{|\psi_{i}\rangle,i=1,\ldots,\chi\}$, with $\chi$ the number of basis, the excited state takes the form $|\Psi\rangle=\sum_ic_i|\psi_i\rangle$, with $c_i$ the variational parameters. To determine $c_i$, one can now diagonalize the Hamiltonian within this subspace, leading to the generalized eigenvalue equation: $\mathbf{H}v=E\mathbf{N}v$, where $v=(c_1,c_2,\ldots,c_{\chi})^{\mathrm{T}}$, and $\mathbf{H}_{ij} = \langle\psi_{i}|H|\psi_{j}\rangle$, $\mathbf{N}_{ij} = \langle\psi_{i}|\psi_{j}\rangle$ are the Hamiltonian matrix and overlap matrix in this subspace, respectively. Note that, the dimension $\chi$ of a fixed $\bm{K}$ sector scales linearly with circuit depth, and is independent of system size, suggesting a scalability in our approach.

On quantum devices, above matrix elements can be obtained through the Hadamard test followed by classical post processing~\cite{Chen2025a}, a specific version of which has been implemented in a recent experiment~\cite{Will2025}. In this approach, for two states $|\psi_i\rangle=u_i|0\rangle^{\otimes},i=1,2$ with unitary $u_i$, the state overlap $\langle \psi_1|\psi_2\rangle$ (and related Hamiltonian matrix element) is measured through a controlled unitary with $u_1^{\dagger}u_2$ acting on the system qubits. Therefore, the sequential circuits in previous section for preparing initial states in all three $(\phi_1,\phi_2)$ sectors appear to be necessary, while it is not clear whether it would be feasible or even possible to use Hadamard test to measure state overlap when two states are prepared with different measurement protocols. The latter is the case for states in different $(\phi_1,\phi_2)$ sectors. However, if one only use one ground state for the excitation spectrum, the sequential circuit can be replaced by the projective measurement and error correction for $W_p$. In this case, the accuracy will decrease in certain sectors, see comparative results in Appendix~\ref{app:Numerical_details}. Note that, promising alternatives to Hadamard test have been proposed in Ref.~\cite{Wang2025b}, for which the sequential circuits for initial states are still needed when considering overlap of states in two different sectors.

Using above approach, we have obtained variational excitation spectrum in the zero vortex sector for three different system size with $(L_1,L_2)=(3,2),(3,3),(4,3)$, respectively. Comparing to fermion solution, all three cases have similar behavior, showing that using three ground states can significantly improve the accuracy of excitation spectrum. Without loss of generality, we present the $(L_1,L_2)=(3,3)$ results in Fig.~\ref{fig:Kitaev_zero_vortex}, and leave the other two cases to Appendix~\ref{app:Numerical_details}. For states with relatively low energy, specifically the 100 lowest energy states of each sector, this method can guarantee the maximum energy error is smaller than $1\times 10^{-12}$, while the energy errors of three ground states are smaller than $1\times 10^{-13}$.

For the three sectors containing the ground states (shown in Fig.~\ref{fig:Kitaev_zero_vortex}(b),(c),(d)), in the fermion picture, the excitations are created by occupying even number of Bogoliubov modes on top of the ground state. Comparing with fermion solution, the results suggest a wide range of excited states can be captured, which can be labeled by the occupied fermion modes $n$ above the ground states. It turns out that for relatively small system size, e.g., $(L_1,L_2)=(3,3)$, using three ground states we can obtain both $n=2$ and $n=4$ states, while only the former is covered when using only one ground state. However, with increasing system size, e.g., $(L_1,L_2)=(4,3)$, only qualitative improvement is observed and the $n=4$ states still have large error. This suggests that multi-particle states (in the spin language) are beyond the scope of this ansatz. Indeed, the ansatz is developed on the single quasi-particle picture, and is not expected to capture multi-particle states. We leave more comparative data to the Appendix~\ref{app:Numerical_details}.

The excitation spectrum in $(-1,-1)$ sector (shown in Fig.~\ref{fig:Kitaev_zero_vortex}(e)) is particularly interesting. From exact solution one can find that the lowest energy state in this sector corresponds to the single fermion excitation~\cite{Lahtinen2008}, giving rise to the fermion gap of this model. This is allowed due to the constraints of the 1-form symmetry (see Appendix~\ref{app:Majorana} for details). Through comparing results using one ground state (see Appendix~\ref{app:Numerical_details}), we find using three ground states significantly improves the accuracy, giving rise to results with low error, while results with any of the three ground states would result in significant error (with error of lowest energy state ranges from $5\times 10^{-4}$ to $3\times 10^{-2}$, depending on which ground state is used). Similar behavior is also observed on larger system size, see Appendix~\ref{app:Numerical_details}. Further comparing to the results using PEPS on the infinite plane, where capturing this single fermion excitation quantitatively or qualitatively seems to be hard~\cite{Chen2025c}, our results on finite torus allow identifying the nature of each low-energy state, and form a complementary approach to the widely used infinite PEPS based excitation ansatz~\cite{Vanderstraeten2019a}.

\subsection{Nonzero vortex sector}
\label{subsec:nonzero_vortex}

We now move onto the non-zero vortex sector. In this case, the vortex would carry a fermion to form a non-abelian anyon, i.e., Ising anyon~\cite{Kitaev2006,Lahtinen2008}. Since different vortex configurations are orthogonal to each other, we consider fixed vortex configurations in this subsection. As our starting point is the ground state, we will focus on the case where the number of vortices is small so that the state has a low energy. 
We further restrict to the case where the vortices are close to each other, so that the picture is close to a one particle excitation and amenable to our excitation ansatz. We leave the other cases to future works.

\begin{figure}[htb]
\centering
\includegraphics[width=0.95\columnwidth]{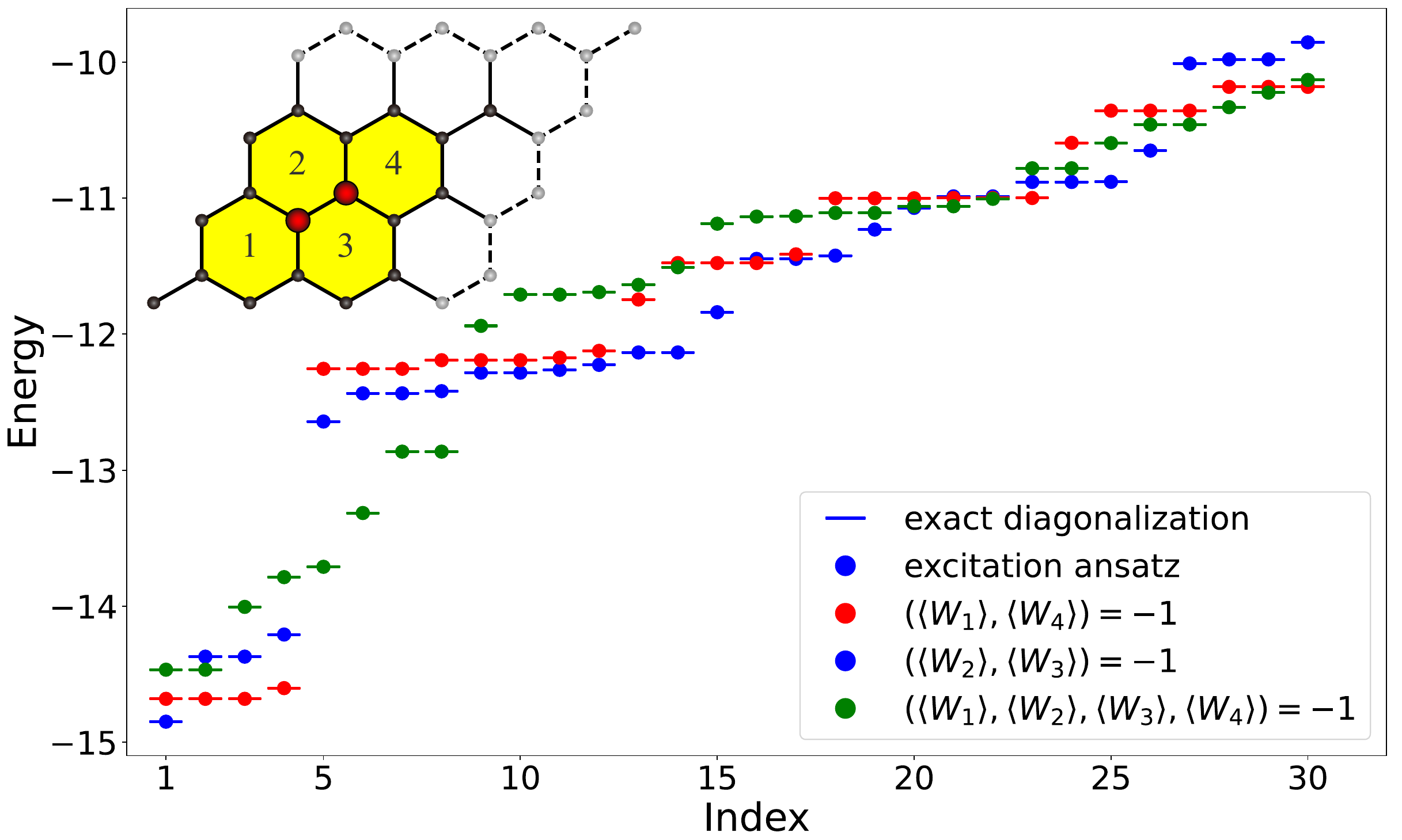}
\caption{
Variational excitation spectrum in nonzero vortex sectors, in comparison with ED. The system size is $(L_1,L_2)=(3,3)$. We consider creating vortices by acting on the red dots shown in the inset.
Three spectra of corresponding $W_p$ configurations are plotted with different colors. The other vortex configurations with $(\langle{W_1\rangle}, \langle{W_2}\rangle)=-1$, $(\langle{W_2\rangle}, \langle{W_4}\rangle)=-1$, $(\langle{W_1}\rangle, \langle{W_3}\rangle)=-1$ and $(\langle{W_3}\rangle, \langle{W_4}\rangle)=-1$ have identical low-energy spectrum as $(\langle{W_2}\rangle, \langle{W_3}\rangle)=-1$.
}
\label{fig:Kitaev_nonzero_vortex}
\end{figure}

With these considerations, we now present the results for $(L_1,L_2)=(3,3)$ with the possible vortex configurations presented in Fig.~\ref{fig:Kitaev_nonzero_vortex}. Related results for other sizes are shown in Appendix~\ref{app:Numerical_details}. Without loss of generality, we consider the four plaquettes around one $x$-link (see Fig.~\ref{fig:Kitaev_nonzero_vortex} for the label). Since the vortex excitations appear in pairs, there are seven types of vortex configurations on the four plaquettes in total, besides the vortex-free configuration we discussed previously. Each of the configurations corresponds to two different two-site Pauli operators acting on the $x$-link respectively, which we will denote as $V$ in the following. Specifically, for the three representative configurations we presented in Fig.~\ref{fig:Kitaev_nonzero_vortex}, 
denoting the site on the highlighted link in $A(B)$ sublattice as $a (b)$
and using the commutation relation between Pauli matrices and plaquette operators $W_p$, it is easy to see that operators $V=\sigma_{a}^{x},\sigma_{b}^{x}$ create vortices on plaquettes 2 and 3, operators $V=\sigma_{a}^{y}\sigma_{b}^{z},\sigma_{a}^{z}\sigma_{b}^{y}$ create vortices on plaquettes 1 and 4, and operators $V=\sigma_{a}^{y}\sigma_{b}^{y},\sigma_{a}^{z}\sigma_{b}^{z}$ create vortices on all plaquettes 1, 2, 3 and 4.

With the vortex configuration fixed, we now discuss how to construct the basis state. Using the relation between Wilson loops and plaquette operators Eq.~\eqref{eq:1_form_symmetry}, one can find that the presence of vortex will alter the eigenvalues of Wilson loops near the vortex, which is a manifestation of broken translation invariance. In principle one can divide the Hilbert space of a given vortex configuration by a finer label with loop dependent $(\phi_1,\phi_2)$. For our purpose, we find it is convenient to only consider vortex label without the label given by Wilson loops, and will take this strategy hereafter.

Since the Wilson loop configurations are not uniform in a vortex configuration, we now consider all three ground states with four possible actions of the loop operators $U_a(a=1,2)$: the presence or absence of $U_1$ or $U_2$. We find that, to achieve a high accuracy with the variational spectrum, it is better to put both the (possible) loop operators $U_a(a=1,2)$ and the two-site operator $V$ on the last layer. And we can further insert local Hamiltonian terms $O$ in the bulk of the circuit, as done in the vortex free case. To reduce error, the loop $U_a$ acting on should contain the link where the local perturbation $O$ acts. A typical basis is then given by:
\begin{equation}
    |\psi\rangle=VU_1U(\bm{\theta}_2)OU(\bm{\theta}_1)|\Psi_0\rangle,
\label{eq:Kitaev_nonzero_vortex}
\end{equation}
where a major difference with Eq.~\eqref{eq:Kitaev_zero_vortex} is the absence of translation operator. Enumerating all basis, one can find that the dimension of the subspace still scales linearly with both system size and circuit depth. Intuitively, one can understand that the three elements play different roles in the basis construction: the link operator $V$ creates the vortex configuration, the loop operator adjusts the fermion boundary conditions, and the Hamiltonian term $O$ creates fermion excitations on top of the ground state.

It turns out some of the vortex configurations share the same low-energy excitation spectrum, and in total there are three different spectrum, shown in Fig.~\ref{fig:Kitaev_nonzero_vortex}. Comparing with ED, we find that the apparently continuous spectrum can be captured to a high accuracy with our approach, where the energy error is smaller than $10^{-12}$ for all vortex configurations we considered. The characteristic quantity, i.e., the vortex gap, can be read out straightforwardly from our symmetry aware variational spectrum.

Therefore, in the nonzero vortex sector, our algorithm performs at the same level of accuracy as in the vortex-free sector, which provides a highly precise spectrum for identification of chiral spin liquid. Together with the zero vortex excitations, the results firmly suggest that our approach can efficiently describe low energy excitations of topological chiral spin liquid phase. As the Kitaev honeycomb model is exactly solvable, one may wonder whether our approach will still provide useful information when the 1-form symmetry is emergent instead of being exact on the lattice. To investigate this question, in the following we will study an abelian $\mathrm{SU}(2)_1$ chiral spin liquid model on the square lattice.

\section{Application to the chiral Heisenberg antiferromagnet on torus}
\label{sec:Square_CSL_torus}

\subsection{The model}
\label{subsec:Square_model}

The model we consider for the $\mathrm{SU}(2)_1$ CSL with emergent 1-form symmetry is defined on the square lattice~\cite{Nielsen2013,Poilblanc2017b}, with a spin-1/2 degree of freedom on each site. The Hamiltonian is given by:
\begin{equation}
\begin{split}
    H_{\mathrm{square}} = &J_{1} \sum_{\langle i,j\rangle}\vec{S}_{i}\cdot \vec{S}_{j} + J_{2}\sum_{\langle\langle i,j\rangle\rangle}\vec{S}_i\cdot \vec{S}_j \\
    &+ i\lambda\sum_{\langle i,j,k,l\rangle_{p}}\left(P_{ijkl} - P_{ijkl}^{-1}\right),
\end{split}
\label{eq:Square_model}
\end{equation}
where following the convention we have used the spin-$1/2$ operators with $\vec{S}=\vec{\sigma}/2$. Here the $J_1$ $(J_2)$ term runs over all (next) nearest-neighbor pairs. $P_{ijkl}$ is the cyclic permutation operator on each plaquette, running, say, clockwise, with the defining relation $P_{ijkl}|\alpha\rangle_i|\beta\rangle_j|\gamma\rangle_k|\delta\rangle_l=|\delta\rangle_i|\alpha\rangle_j|\beta\rangle_k|\gamma\rangle_l$, and can be decomposed into three-spin interactions acting on every triangle in the plaquette~\cite{Poilblanc2017b}. The phase diagram of this model has been studied in Ref.~\cite{Nielsen2013}, and a suitable set of parameters for $\mathrm{SU}(2)_1$ CSL has been proposed and studied via ED and tensor network methods~\cite{Nielsen2013,Poilblanc2017b,Hasik2022}, which we take here: $J_{1} = 2\cos(0.06\pi)\cos(0.14\pi)$, $J_{2} = 2\cos(0.06\pi)\sin(0.14\pi)$, and $\lambda = \sin(0.06\pi)$.

For a generic CSL phase, it is known that on clusters with torus geometry the ground state would be topologically degenerate, which is separated by a gap from the low-energy excitations, akin to the Kitaev honeycomb model studied in previous sections. In contrast, on clusters with open disk geometry, the ground state and the low-energy excitations would form a chiral and linear dispersing branch, whose level content is described by a (1+1)D CFT~\cite{Francesco1997}. Here we aim to find these spectroscopic signatures on quantum circuits, and will focus on the torus case first. 

\subsection{HVA for ground states on torus}
\label{subsec:Square_torus_HVA}

\begin{figure}[hptb]
\centering
\includegraphics[width=0.95\columnwidth]{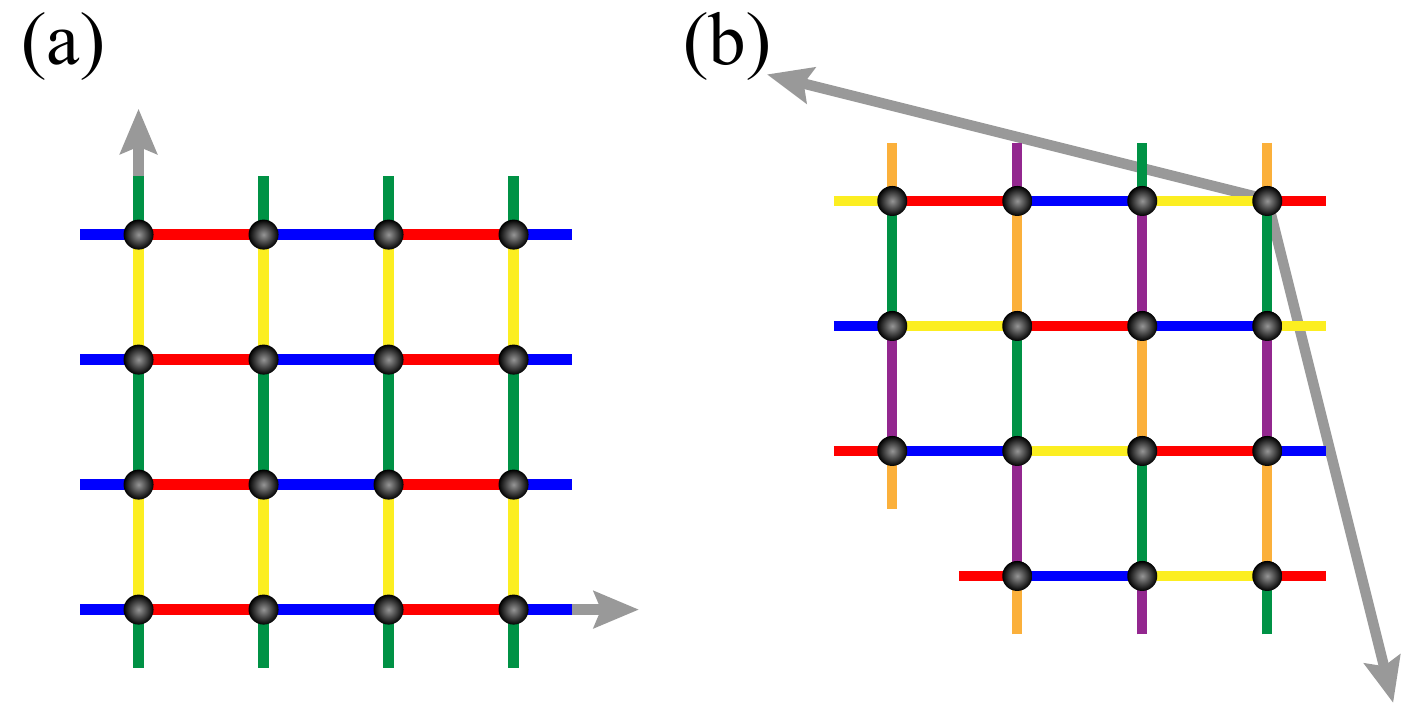}
\caption{Schematics of HVA for the $\mathrm{SU}(2)_1$ CSL on torus geometry. (a) $N_s=16$ site cluster, with torus generated by translation along vectors $\mathbf{t_1}=(4,0), \mathbf{t_2}=(0,4)$. (b) $N_s=15$ site cluster, with torus generated by vectors $\mathbf{t_1}=(-4,1), \mathbf{t_2}=(1,-4)$. In both (a) and (b) links with the same color correspond to gates in one layer.
}
\label{fig:Square_torus_HVA}
\end{figure}

Studying the model on torus geometry can reveal both the topological degeneracy and quasi-particle content of the chiral topological order~\cite{Chen2020,Chen2021}, as we have already seen in the case of Kitaev CSL. The ground state manifold of $\mathrm{SU}(2)_1$ CSL is known to depend on the system size being even or odd. For the even case, the ground state is two-fold topologically degenerate. For the odd case, the degeneracy of ground state manifold is proportional to system size, which is explained by a single anyon localized on arbitrary site in the ground state~\cite{Chen2021}. Here we will take the systems with $N_s=16$ and $N_s=15$ site for each case, with clusters shown in Fig.~\ref{fig:Square_torus_HVA}.

Different from the Kitaev honeycomb model, here the topologically degenerate ground states cannot all be individually and easily addressed via VQE, due to a lack of exact expression of the 1-form symmetry. Thus we will target only part of the ground state manifold first, with the informed physical picture mentioned above. The full content of the ground state manifold will be revealed by the simulations of excitation spectrum on quantum circuits.

To approach the ground state, we use the HVA circuit, in line with the strategy for Kitaev honeycomb model. Here we only consider gates generated by nearest-neighbor Heisenberg interaction, motivated by a recent work on preparing the CSL phase using Floquet engineering~\cite{Mambrini2024}. The ansatz takes a form:
\begin{equation}
    U(\bm{\theta}) = \prod_{i=1}^{D}\prod_{\langle j,k\rangle\in\mathcal{B}_{i}} \mathrm{exp}\left(-i\theta_{i}^{\langle j,k\rangle}\vec{S}_{j}\cdot\vec{S}_{k}\right),
\label{eq:Square_torus_HVA}
\end{equation}
where $D$ is the circuit depth, and $\mathcal{B}_{i}$ is a set containing the non-overlapping gates in the $i$-th layer, as shown by the color in Fig.~\ref{fig:Square_torus_HVA}. We further require that all $\mathcal{B}_{i}$ appear in a periodic sequence with periodicity equals to 4 (6) for $N_s=16 (15)$, so that the circuit is reminiscent of a first order Trotterized time evolution with a nearest-neighbor Hamiltonian. $\bm{\theta}=\{\theta_{1}^{\langle j,k\rangle}, \dots, \theta_{D}^{\langle j,k\rangle}\}$ are the variational parameters of the ansatz, with $\langle j,k\rangle$ runs over all nearest-neighbor links. Due to the absence of one-site translation symmetry in the circuit, the parameter of each gate is taken to be independent.

For the $N_s=16$ site torus, although the ground state is topologically two-fold degenerate, this degeneracy is lifted on finite torus, as revealed by ED results. Thus we target only one ground state in this case. As the ground state is known to be a spin singlet, we organize the neighboring spins into a spin singlet, thus forming a direct product of spin singlets for the initial state. (The actual pairing form is not important here, as long as only one singlet configuration is considered.)
With a circuit depth $D=100$, we obtain variational energy with error smaller than $5\times 10^{-2}$ compared to the ED results. In addition, we have also tried the subspace search VQE algorithm~\cite{Nakanishi2019} for the two ground states, which turns out to have larger error than HVA (data not shown).

In contrast, the ground state degeneracy is much larger on the $N_s=15$ site torus. Here, the ground state not only has a $2\times 4$-fold exact degeneracy (as shown by ED), but also has a topological degeneracy originating from the anyon content of the $\mathrm{SU}(2)_1$ CSL~\cite{Chen2021}. To tackle this situation, we will target $2\times 15$ different ground states, where each one has a different initial state, while using the same type of HVA circuit. The initial state is chosen by isolating one of the spins while pairing the other neighboring spins into singlets, resulting in $15$ configurations. 
The isolated spin can be set in $|0\rangle$ or $|1\rangle$, thus in total $30$ configurations. Since this factor of $2$ comes from $\mathrm{SU}(2)$ symmetry in the Hamiltonian, the unitary gates can be taken to be the same. We then set the isolated spin to state $|0\rangle$, and optimize each of the 15 configurations with different isolated spin separately. The optimized circuit can be applied to the same configuration with isolated spin in $|1\rangle$, resulting in a new variational ground state. With a circuit depth $D=30$, 
the energy error of the lowest energy state is around $0.35$ compared to ED. Note that, the obtained $30$ states are not mutually orthogonal, and are intended for the excitation spectrum presented next.

Note that, in both $N_s=16$ and $N_s=15$ cases, the initial state is close to a product state. Thus it is not surprising that a large circuit depth is needed to transform it into a faithful variational ground state. This is different from the approach in Ref.~\cite{Mambrini2024}, where the highly entangled nearest-neighbor resonating valence bond state~\cite{Fazekas1974, Schuch2012} is used as the initial state, and circuit parameters are fixed according to the Floquet engineering framework.

\subsection{Variational excitation spectrum on torus}
\label{subsec:Square_torus_excitation}

Now we are ready to calculate the excitation spectrum of this model on torus. The procedure follows exactly the way we did for the Kitaev honeycomb model. Here we consider a two-site perturbation acting on each of the nearest-neighbor pair of qubits, with the operator taken from the two-qubit operator basis. The operator can further be inserted in arbitrary circuit depth, resulting in a subspace for excitations with dimension linear in system size $N_s$ and circuit depth. Diagonalizing the Hamiltonian in this subspace, one can find the low energy spectrum of the model. Note that, since the circuits does not have translation symmetry explicitly, we have treated all momentum sectors in one subspace. Nevertheless, with high accuracy for the ground state, momentum quantum numbers can be extracted afterwards by measuring the expectation value of lattice translation operator.

\begin{figure}[hptb]
\centering
\includegraphics[width=0.95\columnwidth]{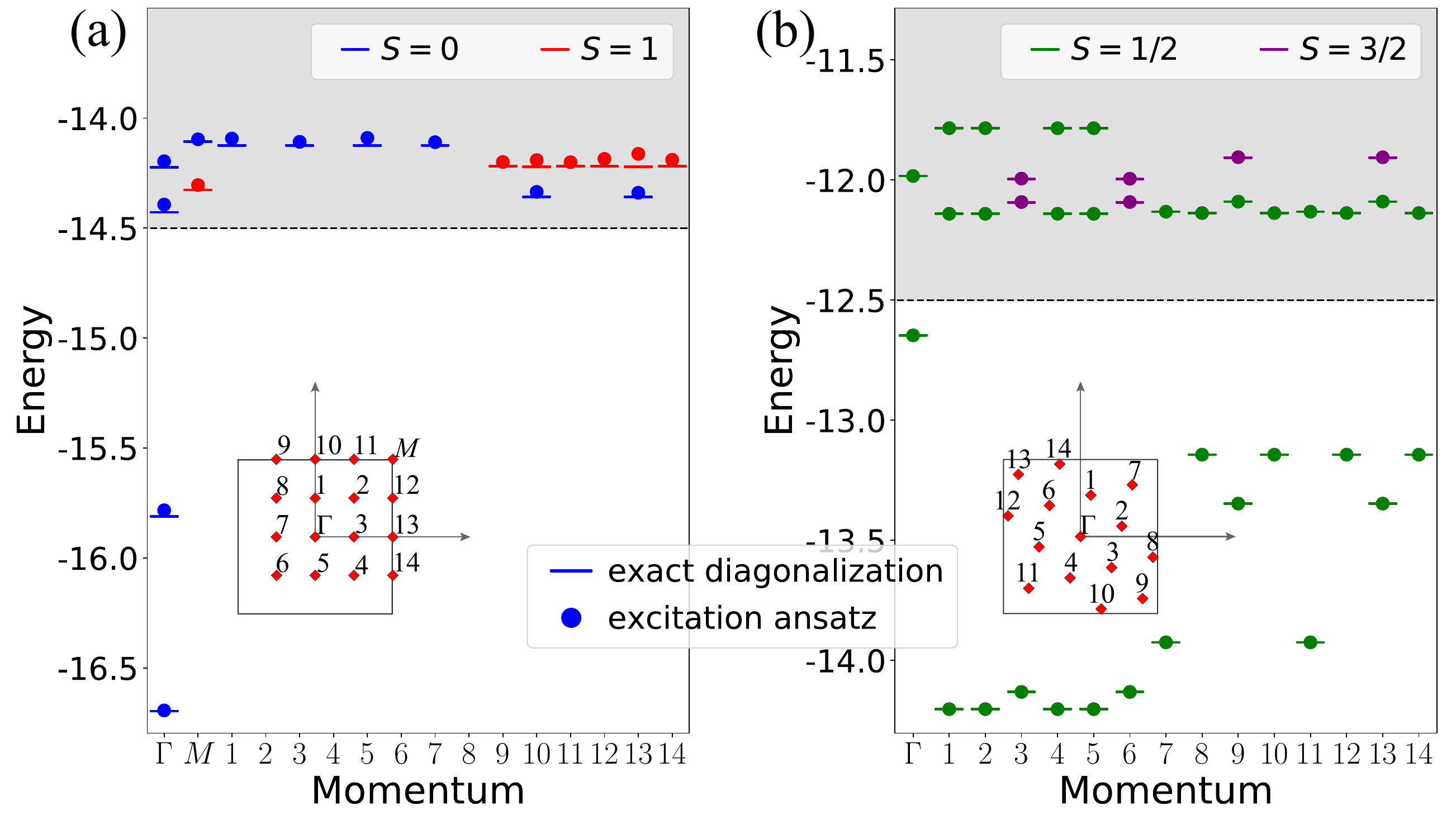}
\caption{Variational low-energy spectrum on torus, in comparison with ED.
(a) $N_s=16$ site. (b) $N_s=15$ site. In both cases, we have used momentum and total spin quantum numbers to label the eigenstates, with the Brillouin zone shown in the inset. Shaded areas in (a) and (b) differentiate ground state manifold from excitations.
}
\label{fig:Square_torus_excitation}
\end{figure}

For the $N_s=16$ case, starting with one ground state, we are able to find the topologically degenerate ground state and a few higher energy excited state, as shown in Fig~\ref{fig:Square_torus_excitation}(a). This is remarkable, demonstrating that our approach provides a way to detect topological degeneracy in a black-box manner. Note that, contrary to the Kitaev honeycomb model where topological degeneracy can be labeled by quantum numbers explicitly, here this degeneracy is emergent, and should be better found from the excitations. Moreover, on top of the ground state manifold, the low-energy excited state are composed by anyon pairs and forms a continuum, which again is captured in a reasonable accuracy by our approach. 

For the $N_s=15$ case, 
on top of the optimized $30$ (non-orthogonal) variational ground states, we construct the tangent space excitations, using only the top $6$ layers to insert perturbation (to reduce computational cost). The result is shown in Fig.~\ref{fig:Square_torus_excitation}(b), demonstrating that even with a highly degenerate ground state manifold, our approach can find the excitations in a reasonable accuracy. More prominently, all the degenerate ground states corresponding to localized anyons can be found using this approach, with a energy error $10^{-2}$ compared to ED. From the Fig.~\ref{fig:Square_torus_excitation}(b), one can find that these degenerate ground states are clearly separated from the rest of the spectrum.

The key observation from the results on torus is the following. With polynomial computational cost, our approach allows to identify emergent topological degeneracy and provides a way to identify signatures of topological order from first principle VQE calculations on quantum circuits. Combining the messages from benchmarks with Kitaev honeycomb model, we have shown that our approach works for both exact and emergent 1-form symmetry.
In the next section, we will show that putting the system on open disk geometry, our approach can in fact provide more evidence for chiral topological order.

\section{Application to a chiral Heisenberg antiferromagnet on open disk}
\label{sec:Square_CSL_disk}

\subsection{HVA for ground state on open disk}
\label{subsec:Square_disk_HVA}

\begin{figure}[hptb]
\centering
\includegraphics[width=0.98\columnwidth]{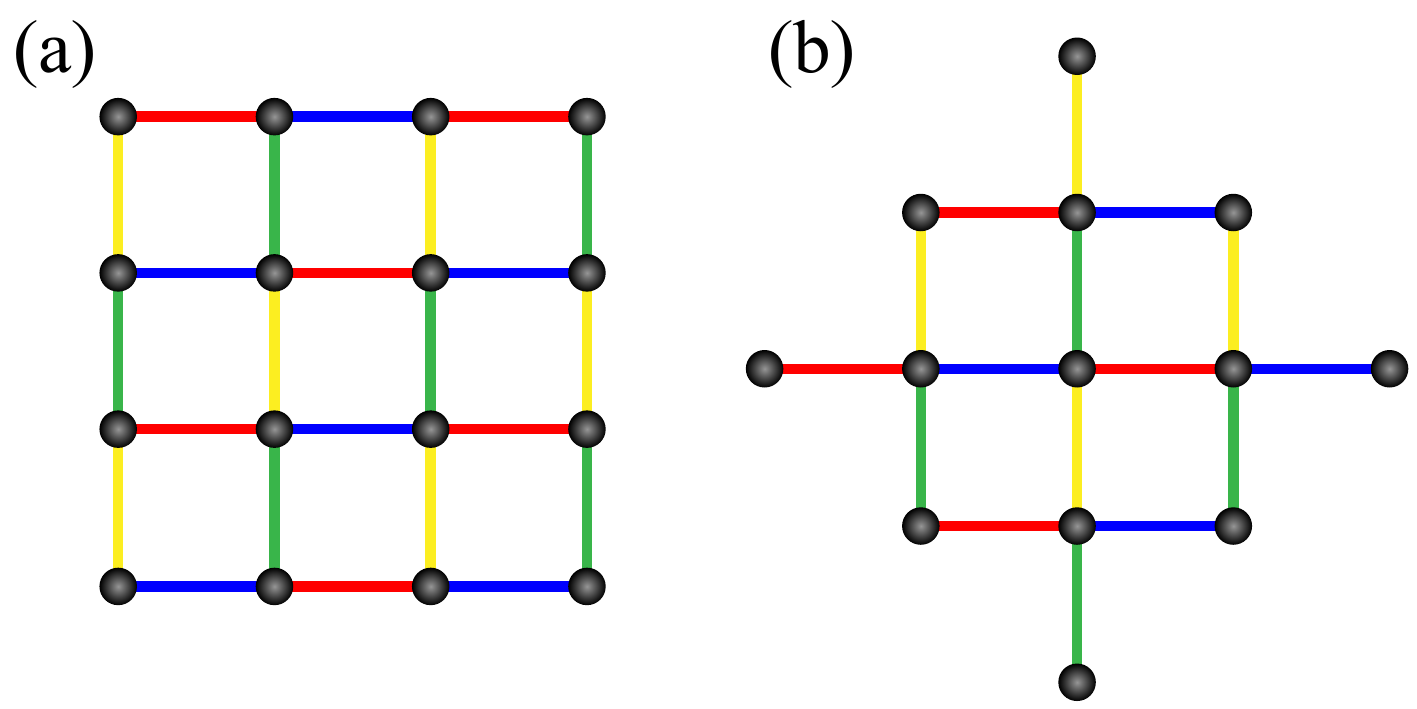}
\caption{Schematics of HVA for $\mathrm{SU}(2)_1$ CSL on open disk geometry. (a) $N_s=16$ site. (b) $N_s=13$ site. In both (a) and (b), links with the same color correspond to gates in one layer.
}
\label{fig:Square_disk_HVA}
\end{figure}

In addition to the torus geometry, studying the model on open disk geometry can reveal edge properties of the CSL, which, under the bulk-edge correspondence, is a hallmark of chiral topological order. For the $\mathrm{SU}(2)_1$ CSL studied here, the relevant CFT is the $\mathrm{SU}(2)_1$ Wess-Zumino-Witten (WZW) model with central charge $c=1$~\cite{Francesco1997}. Since the $\mathrm{SU}(2)_1$ CFT has two sectors, following the widely used strategy in ED study of CSL (see, e.g., Ref.~\cite{Chen2021}), we will consider two different cluster sizes to reveal the CFT spectrum in both sectors. Specifically, we use both 16-site and 13-site clusters with $C_4$ lattice rotation symmetry, shown in Fig.~\ref{fig:Square_disk_HVA}. Note that there are incomplete squares at the edge of the disk in the 13-site model, for which we decompose the plaquette terms in Eq.~\eqref{eq:Square_model} into three-spin interactions and only retain relevant ones for open boundary condition.

Following the procedure for the torus case, we start by optimizing the variational ground state using the HVA, which again requires a proper initial state. For the $N_s=16$ site system, the ground state is a total spin singlet, and the initial state we use is constructed by pairing neighboring spins into a singlet, similar to the torus case with $N_s=16$. In contrast, the ground state of the $N_s=13$ site system is two-fold degenerate, forming a total spin-$1/2$ doublet. Similar to the torus case, suitable initial states are defined in the following way: as shown in Fig.~\ref{fig:Square_disk_HVA}(b), we set the (isolated) spin in the center to $|0\rangle$ and $|1\rangle$, respectively, and all other spins are paired up into short-range singlets.

To find the variational ground state, we again use the HVA generated by nearest-neighbor Heisenberg term in the Hamiltonian Eq.~\eqref{eq:Square_model}. Due to the absence of lattice symmetry in the ground state and the circuit, the parameter of each gate has taken to be independent. For the $N_s=13$ case, we find that although there are two different initial states, we only need to optimize Eq.~\eqref{eq:Square_torus_HVA} for one initial state, and apply the optimized circuit to the other initial state would readily generate the degenerate ground state due to $\mathrm{SU}(2)$ symmetry. The reason for this is that both Hamiltonian and the unitary $U(\bm{\theta})$ in Eq.~\eqref{eq:Square_torus_HVA} are $\mathrm{SU}(2)$ invariant and the two initial states are connected by a global $\mathrm{SU}(2)$ rotation. Using the HVA with depth $D=120$ for the $N_s=16$ open disk and depth $D=100$ for the $N_s=13$ case, we find that the variational ground state energy in both cases is less than $10^{-3}$, compared to ED. The relatively large circuit depth could be due to the relatively complicated form of the Hamiltonian comparing to the Kitaev honeycomb model.

\subsection{Variational excitation spectrum on open disk}
\label{subsec:Square_disk_excitation}

With the optimized circuit for ground state available, we can now compute the excitation spectrum using the tangent space excitation ansatz, in the same way as we did for the torus case. Here the circuit does not have $C_4$ symmetry explicitly, so that we have treated all basis states in one space and the angular momentum quantum number is computed afterwards. In practice, this quantum number can be extracted by measuring the expectation value of lattice rotation operator.

\begin{figure}[hptb]
\includegraphics[width=0.95\columnwidth]{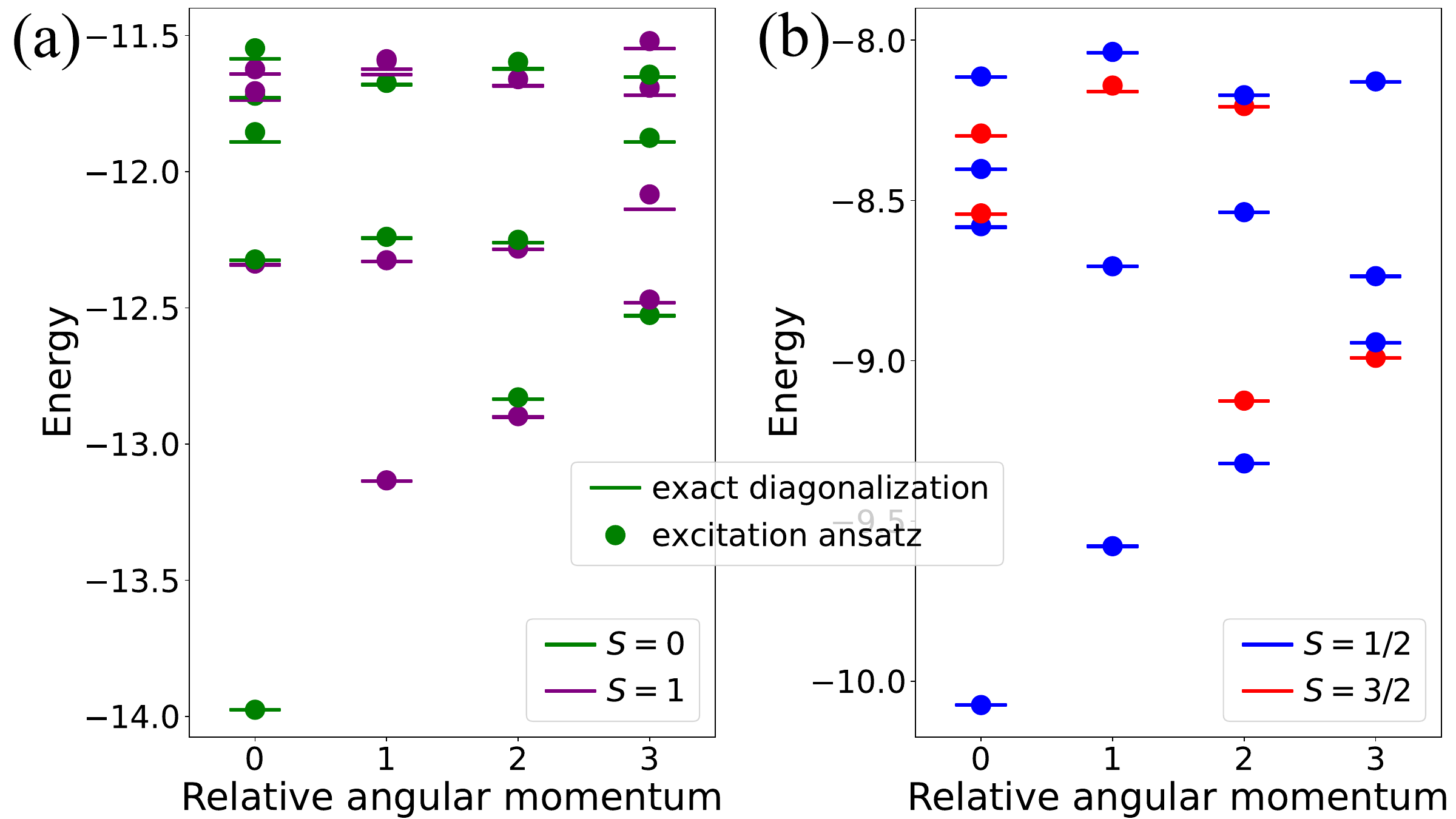}
\caption{
Low-energy spectrum of the $\mathrm{SU}(2)_1$ CSL on open disk. We use angular momentum and $\mathrm{SU}(2)$ spin quantum numbers to label energy levels, with ground state angular momentum subtracted. (a) On the disk with $N_s=16$, the low-energy spectrum follows the $\mathrm{SU}(2)_1$ WZW CFT prediction in the integer sector. (b) On the disk with size $N_s=13$, the low-lying spectrum reveals level counting of the CFT in the half-integer sector. A good agreement between variational result and ED is observed in both sectors.
}
\label{fig:Square_disk_excitation}
\end{figure}

In practice, as the $N_s=13$ case has two degenerated ground states, taking both of them in the basis construction is necessary to obtain excitation spectrum accurately. On the other hand, we also found that for this case one can restrict the perturbation layer to the upper half of the possible depth, which can reduce the computational cost without significantly affecting the low energy spectrum. 
With this strategy, the results for $N_s=16$ and $N_s=13$ are shown in Fig.~\ref{fig:Square_disk_excitation}.

For both system sizes, from the variational spectrum shown in Fig.~\ref{fig:Square_disk_excitation}, one can clearly find a linear dispersing branch in the low-energy spectrum, suggesting chiral edge mode of the CSL phase on open disks. Further checking the $\mathrm{SU}(2)$ quantum number of the states, which are computed afterwards, and comparing the level content with the $\mathrm{SU}(2)_1$ WZW CFT, we find that the low-energy content of the $N_s=16$ case agrees with the integer sector of the CFT, while $N_s=13$ case agrees with the half-integer sector of this CFT. This agreement is manifest for the first three CFT levels, and the variational results further agree with the ED data. Quantitatively, for $N_s=16$, the maximal energy error in the lowest 60 states is smaller than $6\times 10^{-2}$, while ground state energy error is smaller than $5\times 10^{-4}$. For $N_s=13$, we obtain the 60 lowest energy states with the maximum energy error smaller than $2\times 10^{-2}$ and the energy error of two ground states smaller than $5\times 10^{-5}$.

From above results we can see that our method allows to identify smoking gun evidence of CSL phase --- chiral edge mode --- with polynomial computational cost. This is different from existing works where the chiral edge mode is probed by time dependent correlation functions, and entanglement spectrum in tensor network study of CSL. The former is potentially easier to probe on quantum circuits, but does not provide the complete information of the chiral edge theory, while the latter, building on the Li-Haldane conjecture~\cite{Li2008}, requires exponential cost in estimating the reduced density matrix of a sufficiently large subsystem. In contrast, our approach requires polynomial cost and is a native approach on quantum circuits, while providing much more information than correlation functions.

\section{Discussion and outlook}
\label{sec:Outlook}

In this work, we have studied the preparation and characterization of chiral topological phases in quantum circuits. Using the Kitaev honeycomb model and a $\mathrm{SU}(2)_1$ CSL on square lattice as examples, we show, in a constructive way, that chiral state can be prepared in circuits. And for the former case, the preparation is essentially exact. To characterize the prepared chiral state, we provide a way to identify chiral topological phase on quantum circuits via variational energy level spectroscopy. As conventional entanglement measurement is neither feasible nor complete for characterizing chiral phases on quantum circuits, we have resorted to the defining feature of chiral topological order: topological ground state degeneracy and chiral edge mode. We have shown that the tangent space excitation ansatz for quantum circuits can be reliably used for this task, providing a relatively complete characterization of both bulk properties and edge modes.

Concerning both the ground state preparation and the excitations, there are a number of interesting open questions. As the HVA circuit we have used for the variational ground state seems to have a relatively high circuit depth, it would be important to find out whether this ansatz is scalable and if there is a way to reduce circuit depth. Taking insight from the Kitaev honeycomb case, for generic chiral topological phases, it could be beneficial to first prepare a related topological state via measurement and feedforward, on top of which a variational circuit based on HVA could then be applied. As our method can compute the excitations of a chiral model, a natural next step would be to prepare wavepacket of the anyons and study their braiding and fusions. We leave these questions to further studies.

\section*{Acknowledgement}
We thank Yu-Xin Yang for early collaboration on a related topic, and Sylvain Capponi, Xiao-Yu Dong, Roland C. Farrell, Yuhan Liu, Jia-Wei Mei, Georgios Styliaris, Yi Tan, Alexander Wietek, Xiao-Han Yang for discussions.
This work is supported by National Natural Science Foundation of China (Grants No.~12447107, No.~12304186), Guangdong Basic and Applied Basic Research Foundation (Grant No.~2024A1515013065), and Quantum Science and Technology - National Science and Technology Major Project (Grant No.~2021ZD0302100).
D. K. acknowledges the Hakubi projects of RIKEN.

\bibliography{draft}

\newpage
\appendix

\section{Majorana Representation of the Kitaev Honeycomb Model}
\label{app:Majorana}

Denoting $J_x=J_y=J_z=J$ in Eq.~\eqref{eq:Kitaev_model}, we rewrite the isotropic Kitaev honeycomb model in the presence of a weak time-reversal-symmetry-breaking perturbation below:
\begin{align}
	H 
	= J \sum_{\alpha = x,y,z} \sum_{\langle i,j \rangle_{\alpha}} 
	\sigma_{i}^{\alpha} \sigma_{j}^{\alpha} 
	+ K \sum_{\langle i,j,k \rangle \in \triangle} 
	\sigma_{i}^{x} \sigma_{j}^{y} \sigma_{k}^{z},
	\label{Ham_Bphase}
\end{align}
where $\sigma_{j}^{\alpha}$ ($\alpha = x,y,z$) are Pauli matrices and  
$\langle i,j\rangle_{\alpha}$ denotes nearest-neighbor (NN) bonds of type $\alpha$.  
The notation $\langle i,j,k\rangle \in \triangle$ labels ordered triples of sites forming oriented triangles within each hexagonal plaquette. Periodic boundary conditions (PBC) are imposed along both spatial directions.

We adopt a zigzag geometry oriented along the $\bf{a}_2$ direction. For each fixed $i \in \{1,\dots,L_1\}$, the $2L_2$ sites aligned in the $\bf{a}_2$ direction are labeled sequentially by $j = 1,\dots,2L_2$. A single global index is introduced via $\mathcal{I}(i,j) = 2 L_2 (i-1) + j$, for $1 \leq i \leq L_1$ and $1 \leq j \leq 2 L_2$, together with the periodic identifications $\mathcal{I}(\cdot, 2L_2+1) \equiv \mathcal{I}(\cdot,1)$ and $\mathcal{I}(L_1+1,\cdot) = \mathcal{I}(1,\cdot)$. Within each column at fixed $x = i$, sites with odd (even) $j$ are assigned to the $A$ ($B$) sublattice. The total number of sites is therefore $N_s = 2L_{1} L_{2}$.

Under this convention, the three types of NN couplings in Eq.~\eqref{Ham_Bphase} take the form
\begin{align}
	&J \sum_{i = 1}^{L_1} \sum_{j = 1}^{L_2}
	\sigma_{\mathcal{I}(i,2j-1)}^{x}
	\sigma_{\mathcal{I}(i,2j)}^{x},
	\label{eq:x-bonds}
	\\
	&J \sum_{i = 1}^{L_1} \sum_{j = 1}^{L_2}
	\sigma_{\mathcal{I}(i,2j)}^{y}
	\sigma_{\mathcal{I}(i+1,2j-1)}^{y},
	\label{eq:y-bonds}
	\\
	&J \sum_{i = 1}^{L_1} \sum_{j = 1}^{L_2}
	\sigma_{\mathcal{I}(i,2j)}^{z}
	\sigma_{\mathcal{I}(i,2j+1)}^{z},
	\label{eq:z-bonds}
\end{align}
with periodic boundary conditions applied in both spatial directions.
The three-spin terms in Eq.~\eqref{Ham_Bphase} generated by the weak time-reversal-symmetry-breaking perturbation follow the local ordering $\sigma_i^{x}\sigma_j^{y}\sigma_k^{z}$ around each oriented triangle, resulting in
\begin{align}
	K \sum_{i = 1}^{L_1} \sum_{j = 1}^{L_2}
	\Big(
	&\sigma_{\mathcal{I}(i,2j-1)}^{x}
	\sigma_{\mathcal{I}(i,2j)}^{y}
	\sigma_{\mathcal{I}(i,2j+1)}^{z} \notag \\
    &+
    \sigma_{\mathcal{I}(i,2j+2)}^{x}
	\sigma_{\mathcal{I}(i,2j+1)}^{y}
	\sigma_{\mathcal{I}(i,2j)}^{z} \notag 
	\\[2pt]
	&+
	\sigma_{\mathcal{I}(i,2j)}^{x}
    \sigma_{\mathcal{I}(i+1,2j-1)}^{y}
	\sigma_{\mathcal{I}(i,2j+1)}^{z} \notag \\[2pt]
	&+
	\sigma_{\mathcal{I}(i+1,2j+1)}^{x}
	\sigma_{\mathcal{I}(i,2j+2)}^{y}
    \sigma_{\mathcal{I}(i+1,2j)}^{z} \notag \\[2pt]
	&+
	\sigma_{\mathcal{I}(i,2j-1)}^{x}
    \sigma_{\mathcal{I}(i+1,2j-1)}^{y}
	\sigma_{\mathcal{I}(i,2j)}^{z} \notag \\[2pt]
	&+
    \sigma_{\mathcal{I}(i+1,2j)}^{x}
	\sigma_{\mathcal{I}(i,2j)}^{y}
	\sigma_{\mathcal{I}(i+1,2j-1)}^{z}
	\Big).
\end{align}

Each hexagonal plaquette is composed of alternating  $x$-, $y$-, and $z$-type bonds. The associated plaquette operator is given by
\begin{align} 
    W_{p} := &\sigma_{\mathcal{I}(i,2j - 1)}^{z} \sigma_{\mathcal{I}(i,2j)}^{y} \sigma_{\mathcal{I}(i,2j + 1)}^{x} \notag\\
    &\qquad \times \sigma_{\mathcal{I}(i-1,2j+2)}^{z} \sigma_{\mathcal{I}(i-1,2j + 1)}^{y} \sigma_{\mathcal{I}(i-1,2j)}^{x}.
\end{align}
The Wilson loop operators are defined as
\begin{align}
    \Phi_{1,j} &= \prod_{i = 1}^{L_{1}} \sigma_{\mathcal{I}(i,2j - 1)}^{z} \sigma_{\mathcal{I}(i,2j)}^{z}, \\
    \Phi_{2,i} &= \prod_{j = 1}^{2L_{2}} \sigma_{\mathcal{I}(i,j)}^{y}.
\end{align}
For $K = 0$, the Hamiltonian~\eqref{Ham_Bphase} preserves all plaquette operators $W_{p}$ and the ground state resides in the sector satisfying $W_{p} = + 1$ for all $p$. Under PBC, the plaquette operators obey the global constraint $\prod_{p} W_{p} = +1$.

The spin operators can be represented in terms of four Majorana fermions as
\begin{align}
	\sigma_i^{a} = i b_i^{a} c_i, \qquad a \in \{x,y,z\},
\end{align}
where the Majorana operators $\chi_{i} \in \{c_{i}, b_{i}^{x}, b_{i}^{y}, b_{i}^{z}\}$ satisfy $\chi_{i}^{\dagger} = \chi_{i}$ and $\{\chi_{i},\chi_{j}\} = 2 \delta_{ij}$.
To restrict the enlarged Hilbert space to the physical sector, we impose the local constraint
\begin{align}
	D_i := b_i^x b_i^y b_i^z c_{i} = 1.
\end{align}
For each NN bond of $a$-type $\langle i,j\rangle_a$, we introduce a static $\mathbb{Z}_2$ gauge variable
\begin{align}
	u_{ij} = i b_i^{a} b_j^{a},
	\qquad
	u_{ij}^2 = 1,
	\qquad
	[u_{ij},H]=0,
\end{align}
which is conserved on every bond.
The corresponding plaquette flux is given by $W_p = \prod_{\langle i,j\rangle \in p} u_{ij}$.

In terms of the Majorana fermions $\{c_i\}$ and the static gauge variables $\{u_{ij}\}$, the Hamiltonian takes the quadratic form
\begin{align}
	H = -i J \sum_{a=x,y,z} \sum_{\langle i,j\rangle_a} 
	u_{ij}\, c_i c_j
	- iK \sum_{\langle i,j,k\rangle \in \triangle}
	u_{ij} u_{jk}\, c_i c_k.
\end{align}
Treating each $u_{ij}$ as a fixed $\mathbb{Z}_{2}$ constant $\pm 1$, the Hamiltonian becomes purely quadratic in the Majorana operators:  
\begin{align}
	H = \frac{i}{4} \sum_{i,j} A_{ij} c_{i} c_{j},
\end{align}
where $A$ is a real antisymmetric matrix. 
Since any real antisymmetric matrix can be brought into the canonical block-diagonal form, there exists an orthogonal matrix $Q$ such that
\begin{align}
	A = Q \left[\bigoplus_{j = 1}^{L_{1} L_{2}} \begin{pmatrix}
		0 & \varepsilon_{j} \\ -\varepsilon_{j} & 0
	\end{pmatrix}\right] Q^{\mathrm{T}},  
\end{align}
with $\varepsilon_{1} \geq \varepsilon_{2} \geq \dots \geq \varepsilon_{L_{1}L_{2}} \geq 0$. 
Introducing the rotated Majorana operators 
\begin{align}
	\boldsymbol{\eta} =  \begin{pmatrix}
		\eta_{1}^{1} & \eta_{1}^{2} & \dots & \eta_{L_{1}L_{2}}^{1} & \eta_{L_{1}L_{2}}^{2}
	\end{pmatrix}^{\mathrm{T}} := Q^{\mathrm{T}} \mathbf{c},
\end{align}
where $\mathbf{c} = \begin{pmatrix}
c_{1} & c_{2} & \cdots & c_{2L_{1}L_{2}}
\end{pmatrix}^{\mathrm{T}}$, we obtain $2L_{1} L_{2}$ Majorana modes obeying $(\eta_{i}^{\alpha})^{\dagger} = \eta_{i}^{\alpha}$ and $\{\eta_{i}^{\alpha},\eta_{j}^{\beta}\} = 2 \delta_{\alpha \beta} \delta_{ij}$.
In this basis, the Hamiltonian reduces to  
\begin{align}
	H = \frac{i}{2} \sum_{j = 1}^{L_{1} L_{2}} \varepsilon_{j} \eta_{j}^{1} \eta_{j}^{2}.
\end{align}
We now define complex fermion operators for each block via $a_{j} := \frac{\eta_{j}^{1} + i \eta_{j}^{2}}{2}$ and $a_{j}^{\dagger} := \frac{\eta_{j}^{1} - i \eta_{j}^{2}}{2}$ for $j = 1,2, \cdots,L_{1}L_{2}$.  
The product of Majorana operators becomes $\eta_{j}^{1} \eta_{j}^{2} = -i (2a_{j}^{\dagger} a_{j} - 1)$. 
Substituting this back into the Hamiltonian gives the diagonal form 
\begin{align}
	H = \sum_{j = 1}^{L_{1} L_{2}} \varepsilon_{j} \left(a_{j}^{\dagger} a_{j} - \frac{1}{2} \right), 
\end{align}
so that each fermionic mode labeled by $j$ carries an excitation energy of $\varepsilon_{j}$. 


In the Majorana representation, the Hilbert space is enlarged, and physical spin states are obtained by imposing the local constraints $D_{i} = b_{i}^{x} b_{i}^{y} b_{i}^{z} c_{i} = 1$ on every site. For a fixed $\mathbb{Z}_{2}$ gauge background $\{u_{ij}\}$ on a torus, these constraints combine a global projection condition that restricts the fermionic Fock space of the quadratic matter Hamiltonian.
After bringing the antisymmetric Majorana Hamiltonian $A^{u}$ to its canonical block-diagonal form, the matter sector is described in terms of the complex fermionic modes $a_{j}$, and $\hat{N}_{f} = \sum_{j} a_{j}^{\dagger} a_{j}$ denotes the total number of occupied matter-fermion modes.
The corresponding matter-fermion parity operator is $\hat{\pi} = (-1)^{\hat{N}_{f}}$.
As shown in Ref.~\cite{pedrocchi2011physical}, the projection onto the physical subspace fixes the allowed fermion-parity sector in a given gauge background (including the choice of Wilson-loop fluxes) to a definite eigenvalue $\pi_{\mathrm{phys}}(u) = \pm 1$
\begin{align}
    \pi_{\mathrm{phys}}(u) = (-1)^{\theta} \det(Q^{u}) \prod_{\langle ij \rangle} u_{ij},
\end{align}
where $Q^{u}$ is the real orthogonal matrix that diagonalizes $A^{u}$, and $\theta$ is a geometry-dependent constant (for a rectangular torus without twists, $\theta = L_{1} + L_{2}$).
Consequently, a many-body eigenstate $\ket{\Psi}$ of the quadratic Majorana Hamiltonian corresponds to a physical spin state if and only if it satisfies the parity condition $\hat{\pi} \ket{\Psi} = \pi_{\mathrm{phys}}(u) \ket{\Psi}$.
States with the opposite matter-fermion parity are unphysical and must be discarded.
This criterion provides a direct, unambiguous procedure for extracting the physical excitation spectrum within each fixed-flux sector.

For a fixed configuration of the $\mathbb{Z}_{2}$ gauge variables $\{u_{ij}\}$, the quadratic Majorana Hamiltonian defines a spectrum of many-body states subject to the fermion-parity constraint discussed above.
Different gauge configurations may realize the same plaquette-flux pattern $\{W_{p}\}$ while differing by the values of the Wilson-loop operators, or equivalently by the boundary conditions imposed on the Majorana fermions on a torus.
The ground state of the Kitaev honeycomb model is the state with the lowest energy among all physical states.
This procedure is carried out by restricting to the vortex-free sector with $w_{p} = +1$ for all plaquettes and comparing physical states associated with inequivalent $\mathbb{Z}_{2}$ gauge choices within this sector, which are commonly taken to correspond to antiperiodic boundary conditions along the $\mathbf{a}_{1}$ direction and periodic boundary conditions along the $\mathbf{a}_{2}$ direction~\cite{jin2021density,sala2025stability}.

The excitation spectrum in the Majorana representation can be organized into two layers, one associated with matter and the other with gauge degrees of freedom.
For a fixed $\mathbb{Z}_{2}$ gauge background $\{u_{ij}\}$ (and hence fixed plaquette fluxes $\{W_{p}\}$), the quadratic Majorana Hamiltonian is fixed, and its spectrum is obtained by creating matter-fermion excitation, i.e., by occupying higher-energy single-particle eigenmodes, subject to the fermion-parity constraint discussed above. 
In addition, one may consider changing the $\mathbb{Z}_{2}$ gauge background itself: modifying bond variables so as to flip one or more plaquette fluxes $W_{p}$ generates vison (flux) excitations and moves the system to a different flux sector.
In that case, the quadratic Majorana Hamiltonian is changed, and the corresponding matter spectrum must be recomputed in the new gauge background, again with the appropriate physical-parity projection.

The quadratic Majorana Hamiltonian may preserve lattice translation symmetry for a fixed $\mathbb{Z}_{2}$ gauge background $\{u_{ij}\}$, possibly up to a $\mathbb{Z}_{2}$ gauge transformation and boundary-condition twists.
When this condition is satisfied, crystal momentum can be assigned to the matter-fermion eigenmodes and to the resulting many-body states.
Lattice translations act on the Majorana operators by permuting site labels and are represented by real orthogonal operators $P_{x}$ and $P_{y}$, corresponding to translations by one unit cell along the $x$ and $y$ directions, respectively.
On a torus, translations that wrap around the system may acquire a minus sign, reflecting periodic or antiperiodic boundary conditions; this effect can be encoded by twist parameters $\alpha_{x},\alpha_{y} \in \{0,1/2\}$.
In a translation-invariant gauge sector, the Majorana Hamiltonian is invariant under translations up to a gauge transformation, allowing the single-particle eigenmodes to be chosen as simultaneous eigenstates of $P_{x}$ and $P_{y}$. 
The associated crystal momenta are then defined from the phases of the translation eigenvalues and are quantized as $k_{x} = (2 \pi / L_{1})(n_{x} + \alpha_{x})$ and $k_{y} = (2 \pi / L_{2}) (n_{y} + \alpha_{y})$, with integers $n_{x}$, $n_{y}$.
In the presence of energy degeneracies, the eigenmodes are resolved within each degenerate subspace so as to diagonalize the translation operators.
Many-body eigenstates constructed by occupying a set of single-particle modes then carry total crystal momentum given by the sum of the momenta of the occupied modes, reflecting the additive structure of lattice translations.
For comparison with spin spectra, one should consider the crystal momentum of the physical many-body state obtained after imposing the parity constraint, which corresponds to that of the associated spin eigenstate.
It should be emphasized, however, that this momentum assignment is meaningful only in gauge sectors that preserve lattice translation symmetry, possibly up to a $\mathbb{Z}_{2}$ gauge transformation.
For generic gauge configurations, such as those containing localized vortex excitations, translation symmetry is broken and momentum ceases to be a good quantum number.
In this case, the Majorana Hamiltonian no longer commutes with the lattice translation operators, and the single-particle eigenmodes cannot be chosen as simultaneous eigenstates of translations.

\section{Sequential circuits for initial state preparation}
\label{app:ini_state}

To prepare the chiral spin liquid ground state of Eq.~\eqref{eq:Kitaev_model} on quantum circuits, it is useful to start with a suitable initial state, where all plaquette operators $W_p$ and Wilson loop operators $\Phi_1,\Phi_2$ take their eigenvalues~\cite{Kalinowski2023,Sun2023,Bespalova2021}. Indeed, a symmetry aware unitary circuit in VQE is less prone to the local minima and easier to optimize than a fully general circuit. Since the three topologically degenerate ground states all have $\langle W_p\rangle=1$, while the $\Phi_1,\Phi_2$ can take different eigenvalues, in the following we 
consider each case separately. Further, as the three ground states are all translationally invariant with zero momentum, we require that the initial states are also translation invariant.

As all the $W_p$'s can be viewed as local stabilizers, a straightforward way to enforce the $+1$ eigenstate of $W_p$ would be through projection, with projector given by $P=\prod_p (1+W_{p})/2$. On quantum devices, this projection can be realized through measuring the plaquette operator on each hexagon, followed by post-selection or error correction~\cite{Kalinowski2023,Bespalova2021}, which has been realized in a recent experiment~\cite{Evered2025}. The eigenstate of $\Phi_1$ and $\Phi_2$ can be realized by requiring that two spins on each $x$-link form a Bell pair, e.g., $|00\rangle+|11\rangle$. However, to avoid the ambiguous phase factor associated with measurement, later we will find it would be more convenient to use a unitary circuit for the initial state preparation, so that the state with all the good quantum numbers can be obtained deterministically from a product state.

\begin{figure}[hptb]
\centering
\includegraphics[width=0.95\columnwidth]{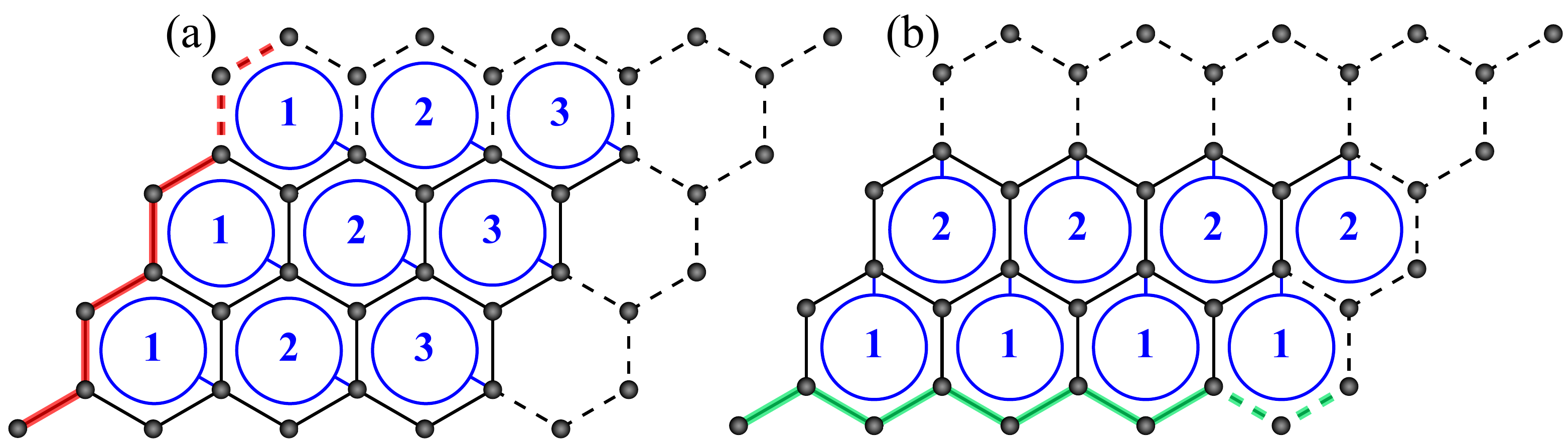}
\caption{Initial state preparation through sequential circuits. In both (a) and (b), the circle represent
a controlled unitary with control qubit indicated by the line segment, which realizes the projection to $\langle W_p\rangle=1$ on the hexagon. The order of the control unitary are indicated by the numbers. 
(a) The qubits on the first column are initialized in an eigenstate of $\Phi_2$ operator with GHZ type entanglement, while the other target qubits are set in the $\sigma^z$ eigenstate. (b) The qubits on the first row are initialized in an $\Phi_1$ eigenstate with GHZ type entanglement, and the other target qubits are in the $\sigma^y$ eigenstate. Depending on the sectors and system size, the initial state can vary.
}
\label{fig:Kitaev_sequential_circuit}
\end{figure}

It is known that sequential quantum circuits provide a suitable framework to prepare non-chiral topologically ordered states on quantum circuits~\cite{Wei2022,Liu2022,Chen2024}. In the same vein, here we find that simple sequential circuits (shown in Fig.~\ref{fig:Kitaev_sequential_circuit}) work for most cases, with the details vary for different eigenvalues of $\Phi_1,\Phi_2$, which further depends on the system size. Thus we will describe the prescriptions separately.

\subsection{The sector with $\phi_1=1,\phi_2=\pm 1$}

The sequential circuit for the $\langle W_p\rangle=1$, $\phi_1 =1$, $\phi_2 =\pm 1$ sectors with arbitrary system size $2L_1L_2$ is shown in Fig.~\ref{fig:Kitaev_sequential_circuit}(a). The procedure starts by assigning a subset of qubits as the control qubits. Using the shaded hexagon in Fig.~\ref{fig:Kitaev_model} as an example, we take qubit-$5$ as the control qubit for this hexagon, whose state is set to $(|0\rangle+|1\rangle)/\sqrt{2}$ (here $|0\rangle,|1\rangle$ are the computational basis states). Then we implement a controlled-$X$ gate on qubit-$1$, i.e.
\begin{equation}
U_{5,1}=|0\rangle\langle 0|_5\otimes\mathbb{I}_1 + |1\rangle\langle 1|_5\otimes\sigma_1^x.
\end{equation}
Similarly, a controlled-$Y,(Z,X,Z)$ gate on qubit-$2,(3,4,6)$ is implemented, respectively, which is followed by a phase gate $S=\mathrm{diag}(1,i)$ on the control qubit. One can find that, the effect of this unitary procedure on the corresponding hexagon is to transform state of the control qubit and arbitrary initial state of the target qubits $\otimes_{i=1,2,3,4,6}|s_i\rangle$ from 
\begin{equation}
|\psi_p\rangle=\otimes_{i=1,2,3,4,6}|s_i\rangle \otimes (|0\rangle+|1\rangle)_5/\sqrt{2}
\end{equation}
into 
\begin{equation}
|\psi_p'\rangle=(1+W_p)\otimes_{i=1,2,3,4,6}|s_i\rangle|0_5\rangle/\sqrt{2},
\label{eq:projection}
\end{equation}
which realizes the projection with $W_p$ through unitary gates.

Above procedure for one hexagon can be implemented simultaneously for all hexagons in one column along $\bf{a}_2$. (Note that, the control qubit needs to be un-entangled with the rest of the qubits before the controlled unitary.) Then we move to the next column, repeating this procedure, until the right most column of unit cells, see Fig.~\ref{fig:Kitaev_sequential_circuit}(a). With this procedure, we have implemented the projection operator on each column except the right boundary ones, which arise due to PBC. For this sequential circuit, only the control qubits have a fixed initial state, while the other qubits can be in arbitrary initial product state. Using this freedom, we now move to the constraints given by Wilson loop operators $\Phi_1,\Phi_2$ and translation invariance, for which we discuss the $L_2$ even and odd cases separately.

For the $L_2$ even case, as the control qubits are in the state $|0\rangle$ in Eq.~\eqref{eq:projection}, to enforce translation along $x$-direction, it is reasonable to initialize qubits of the first column along $\bf{a}_2$ in a Greenberger–Horne–Zeilinger (GHZ) state $(|0\rangle^{\otimes 2L_2}+|1\rangle^{\otimes 2L_2})/\sqrt{2}$, which contains the all zeros configuration for the control qubits and is also an eigenstate of the $\Phi_2$ operator with eigenvalue $+1$. For each of the other target qubits, we set the initial state to $|0\rangle$. Then the state after the sequentially implementing the controlled unitary is given by
\begin{equation}
|\Psi_{1,1}^{\mathrm{even}}\rangle=c\prod_{p}(1+W_{p})(1+\Phi_{2,1})|0\rangle^{\otimes 2L_1L_2},
\label{eq:initial_state_00}
\end{equation}
where we have used $|0\rangle^{\otimes 2L_2}+|1\rangle^{\otimes 2L_2}=(1+\Phi_2)|0\rangle^{\otimes 2L_2}$ for $L_2$ even, with $\Phi_{2,1}$ is the $\Phi_2$ operator supported on the first column, 
$c=2^{-(L_1L_2-L_2+1)/2}$ is a normalization factor and the product runs over all but the right boundary hexagons.

To show that $|\Psi_{1,1}^{\mathrm{even}}\rangle$ is translationally invariant and also a common eigenstate of $W_p$ and $\Phi_1,\Phi_2$ with eigenvalues all equal to $+1$, we note that, using the relation between $\Phi_1$ and $W_p$ (Eq.~\eqref{eq:1_form_symmetry}), one can first find all the $W_p$ have eigenvalue $+1$. Then as the first column has $\Phi_2$ eigenvalue $+1$, one can iteratively find all $\Phi_2$'s have eigenvalues $+1$. Thus using eigenvalue equations of $W_p$ and $\Phi_2$, we can insert appropriate projectors into Eq.~\eqref{eq:initial_state_00} to find 
\begin{equation}\nonumber
|\Psi_{1,1}^{\mathrm{even}}\rangle=c'\prod_{p}(1+W_{p})\prod_{i=1}^{L_1}(1+\Phi_{2,i})|0\rangle^{\otimes 2L_1L_2},
\end{equation}
where $c'$ is a normalization factor and the product now runs over all hexagons and columns. With above expression, the translation invariance along $\bf{a}_1$ and $\bf{a}_2$ with zero momentum now becomes transparent.

For the sector with $\langle W_p\rangle=1$, $\phi_1=1$, $\phi_2=-1$, we can replace the initial GHZ state $(|0\rangle^{\otimes 2L_2}+|1\rangle^{\otimes 2L_2})/\sqrt{2}$ on the first column with another GHZ state $(|0\rangle^{\otimes 2L_2}-|1\rangle^{\otimes 2L_2})/\sqrt{2}$, so that the $\Phi_2$ acting on first column gets the eigenvalue $-1$. With the other parts of circuit taking the same form, the state for this sector is given by:
\begin{equation}
|\Psi_{1,-1}^{\mathrm{even}}\rangle=c\prod_p(1+W_{p})(1-\Phi_{2,1})|0\rangle^{\otimes 2L_1L_2},
\label{eq:initial_state_01}
\end{equation}
where we have used $|0\rangle^{\otimes 2L_2}-|1\rangle^{\otimes 2L_2}=(1-\Phi_2)|0\rangle^{\otimes 2L_2}$. One can verify straightforwardly that this state is a common eigenstate for $W_p$, $\Phi_1$ and $\Phi_2$ with prescribed eigenvalues, and is also translationally invariant in both $\bf{a}_1$ and $\bf{a}_2$ direction with momentum zero.

For the $L_2$ odd case, since $\Phi_2(|0\rangle^{\otimes 2L_2}+|1\rangle^{\otimes 2L_2})=-(|0\rangle^{\otimes 2L_2}+|1\rangle^{\otimes 2L_2})$, one can simply use Eq.~\eqref{eq:initial_state_00} for $\phi_2=-1$ while Eq.~\eqref{eq:initial_state_01} is a suitable state for $\phi_2=1$.

\subsection{The sector with $\phi_1=-1,\phi_2=1$}

For the $\langle W_p\rangle=1$, $\phi_1=-1$, $\phi_2=1$ sector, we find that changing initial product state would work for certain system sizes, while not for all cases.
We now present a different sequential circuit (shown in Fig.~\ref{fig:Kitaev_sequential_circuit}(b)), to make the result fully general for arbitrary $L_1$ and $L_2$.
The control qubits are now given by qubit-3 on each of the hexagon with the state set to $|0\rangle=(|y+\rangle+|y-\rangle)/\sqrt{2}$, where $|y\pm\rangle=\frac{1}{\sqrt{2}}(|0\rangle\pm i|1\rangle)$ is the $\pm 1$ eigenstate of $\sigma^y$ operator. Then in one hexagon we implement a controlled-$X$ operation on qubit-$1$, with the unitary given by:
\begin{equation}
U_{3,1}=|y+\rangle\langle y+|_3\otimes \mathbb{I}_1 + |y-\rangle\langle y-|_3\otimes \sigma^x_1.
\end{equation}
Similarly, a controlled-$Y(X,Y,Z)$ with target qubit-$2(4,5,6)$ is implemented. Notice that $\sigma^z|y+\rangle=|y-\rangle$, thus the effect of this sequence of controlled unitary is to implement a projection for $W_p$ on this hexagon.

Following the case for $(\phi_1,\phi_2)=(+1,\pm1)$ sectors, one can iteratively implement the controlled unitary. With the initial state for the first row being $(|y+\rangle^{\otimes 2L_1}-|y-\rangle^{\otimes 2L_1})/\sqrt{2}$ and each of all other target qubits initialized in $|y+\rangle$, we arrive at the final state:
\begin{equation}
    |\Psi_{-1,1}\rangle=\tilde{c}\prod_p(1+W_{p})(1-\Phi_{1,1})|y+\rangle^{\otimes 2L_1L_2},
\label{eq:initial_state_10}
\end{equation}
where we have used $|y+\rangle^{\otimes 2L_1}-|y-\rangle^{\otimes 2L_1}=(1-\Phi_1)|y+\rangle^{\otimes 2L_1}$ for arbitrary $L_1$, and $\Phi_{1.1}$ is the $\Phi_1$ operator acting on the first row. $\tilde{c}=2^{-(L_1L_2-L_1+1)/2}$ is the normalization factor and the product over $p$ runs on all plaquettes except the top boundary row. One can follow the same reasoning in $(\phi_1,\phi_2)=(+1,\pm1)$ case to show that above state has the right quantum numbers and is translationally invariant with momentum zero.

\subsection{The sector with $\phi_1=-1,\phi_2=-1$}

The sector with $\phi_1=-1$, $\phi_2=-1$ is a bit more involved. For systems with $L_1$ even and $L_2$ odd, we can set the initial state of the first row as $(|y-,y+\rangle^{\otimes L_1}-|y+,y-\rangle^{\otimes L_1})/\sqrt{2}$, with each of the other target qubits in $|y-\rangle$. Following the same circuit as for Eq.~\eqref{eq:initial_state_10}, the resulting state is given by:
\begin{equation}
|\Psi_{-1,-1}^{\mathrm{EO}}\rangle = \tilde{c}\prod_p(1+W_{p})(1-\Phi_{1,1})|y-,y+\rangle^{\otimes L_1L_2}.
\label{eq:initial_state_11_EO}
\end{equation}
For $L_1$ odd and $L_2$ even, the first column should be set to $(|01\rangle^{\otimes L_2}-|10\rangle^{\otimes L_2})/\sqrt{2}$, and all other target qubits are in state $|1\rangle$. The circuit is the same as in the case for Eq.~\eqref{eq:initial_state_00}, with the state given by:
\begin{equation}
|\Psi_{-1,-1}^{\mathrm{OE}}\rangle=c\prod_p(1+W_{p})(1-\Phi_{2,1})|01\rangle^{L_1L_2}.
\label{eq:initial_state_11_OE}
\end{equation}
For the case with $L_1$ and $L_2$ both odd, both Eqs.~\eqref{eq:initial_state_11_EO} and \eqref{eq:initial_state_11_OE} would work.

For the case with both $L_1$ and $L_2$ even, we find it is relatively easy to to prepare a state that is $-1$ eigenstate of both $\Phi_1$ and $\Phi_2$. Indeed, using the loop operators $U_a(a=1,2)$ introduced in the main text, one can change the eigenvalue of Wilson loop operators. However, the resulting state necessarily breaks the translation invariance. To restore the translation symmetry, we can follow the VQE scheme, targeting a translationally invariant state with the desired eigenvalues of Wilson loop operators. Below we illustrate how this works using a minimal example with $(L_1,L_2)=(2,2)$. We have checked that similar scheme works for larger system size.

The target state $|\Psi_{-1,-1}^{\mathrm{EE}}\rangle$ can be prepared by applying a series of projection operators to a translational invariant direct product state. The product state we choose here is $\otimes_{i\in A}|+\rangle_i\otimes_{j\in B}|y+\rangle_j$, where $A,B$ are two sublattices of the system. The operator that project the state onto desired sector is $P=\prod_{S_{i}}(\mathbb{I}+\lambda_{i} S_{i})/2$, where $S_{i}$'s are the local conserved quantities including all $W_{p}$'s, $\Phi_1$ and $\Phi_2$, with eigenvalue $\lambda_i$. In this case, $\lambda_{i}=+1$ for $W_{p}$'s and $\lambda_{i}=-1$ for $\Phi_1$, $\Phi_2$.

\begin{figure}[hptb]
    \centering
    \includegraphics[width=0.9\columnwidth]{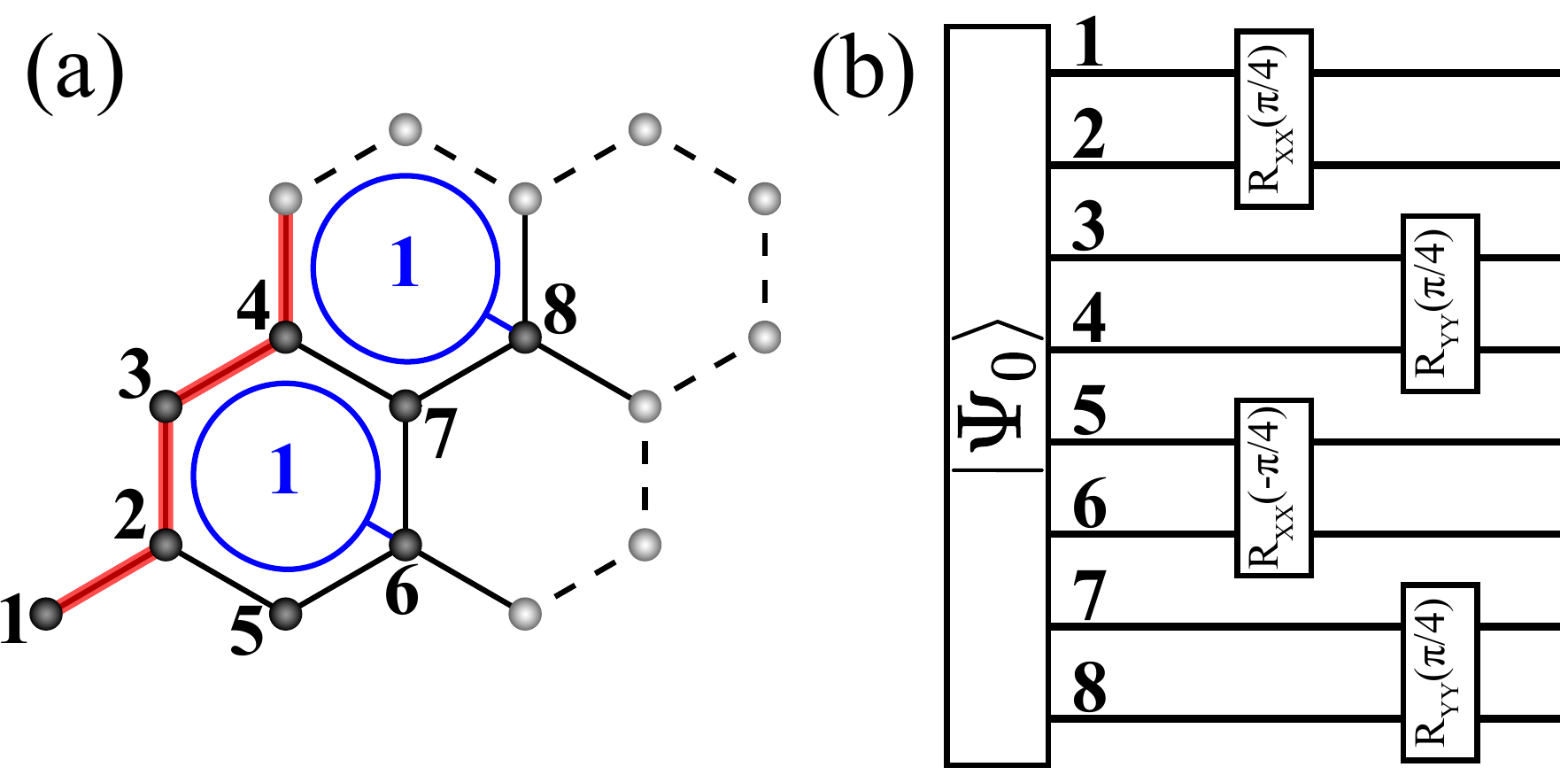}
    \caption{Variational preparation of translationally invariant state $|\Psi_{-1,-1}^{\mathrm{EE}}\rangle$ with $L_1=2,L_2=2$. (a) shows the site label and the initial state. (b) shows the optimized circuit, where $R_{\mathrm{XX}}(\theta)=e^{-i\theta \sigma^x_1\sigma^x_2}$ and $R_{\mathrm{YY}}(\theta)=e^{-i\theta \sigma^y_1\sigma^y_2}$ for qubits-$1,2$.
    }
    \label{fig:Kitaev_initial_variational}
\end{figure}

Starting from the state $|\Psi_0\rangle=U_2|\Psi_{1,-1}^{\mathrm{even}}\rangle$ (cf. Eq.~\eqref{eq:initial_state_01}), we now add local unitary gates on $|\Psi_0\rangle$, and variationally optimize the gate parameters to approach $|\Psi_{-1,-1}^{\mathrm{EE}}\rangle$. The cost function is then given by $\mathcal{C}=1-|\langle\Psi_{-1,-1}^{\mathrm{EE}}|U(\bm{\theta})|\Psi_0\rangle|^2$, where $\bm{\theta}$ are the gate parameters. Since both $|\Psi_0\rangle$ and $|\Psi_{-1,-1}^{\mathrm{EE}}\rangle$ have the right quantum numbers, a reasonable set of gates is generated by the time evolution operator of local Hamiltonian. Through experiments, we found that a two-layer circuit with two-qubit gates is sufficient, shown in Fig.~\ref{fig:Kitaev_initial_variational}.

\section{Additional Numerical results}
\label{app:Numerical_details}

\subsection{Kitaev honeycomb model: Zero vortex sector}

\begin{figure}[htbp]
    \centering
    \includegraphics[width=0.95\columnwidth]{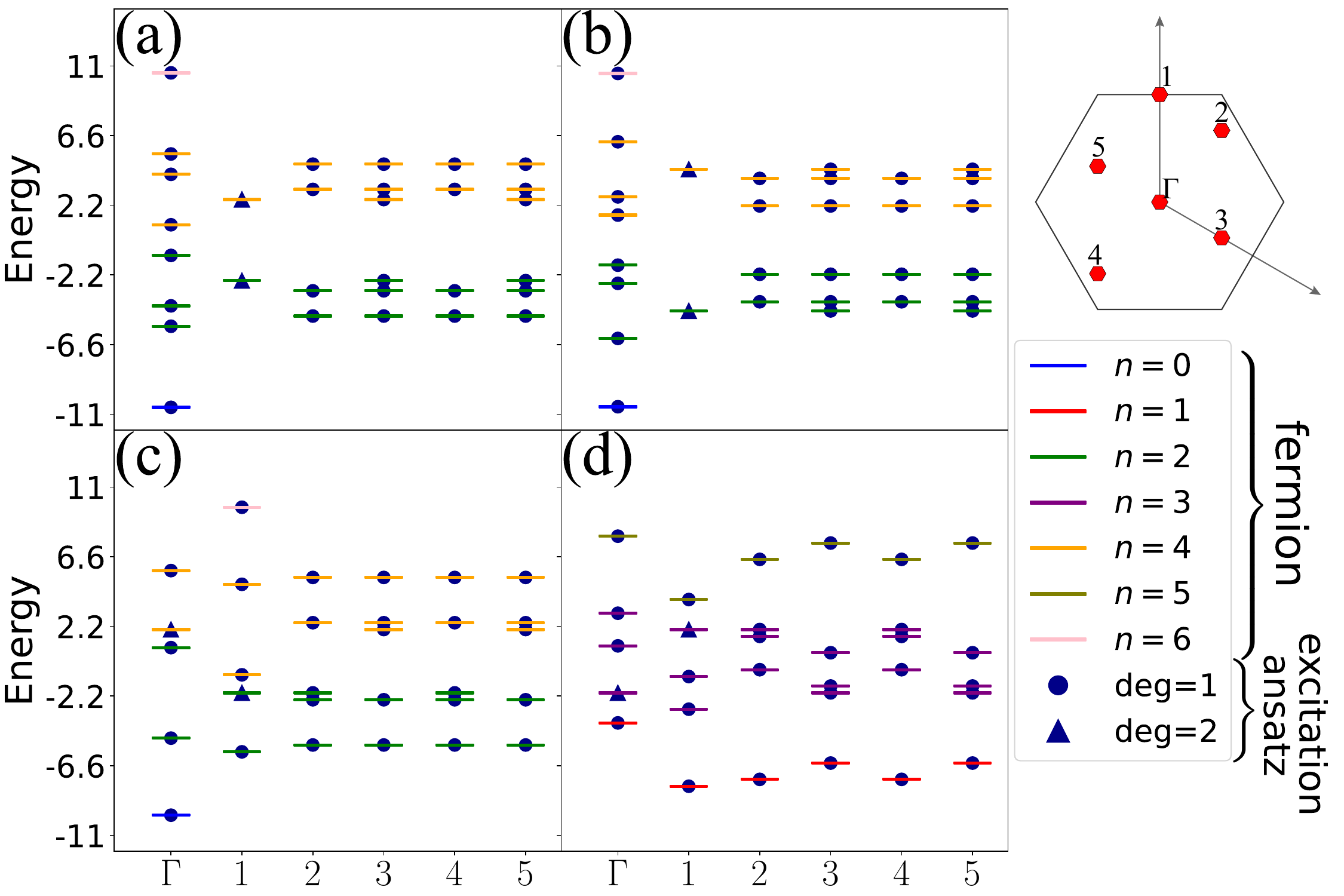}
    \caption{Zero vortex spectrum of a $12$-site torus, comparing with fermion solution. Subplot (a), (b), (c) and (d) correspond to sector $(\phi_1,\phi_2)=(+1,+1), (-1,+1), (+1,-1), (-1,-1)$, respectively.}
    \label{fig:Kitaev_zero_vortex_12site}
\end{figure}

\begin{figure}[htbp]
    \centering
    \includegraphics[width=0.95\columnwidth]{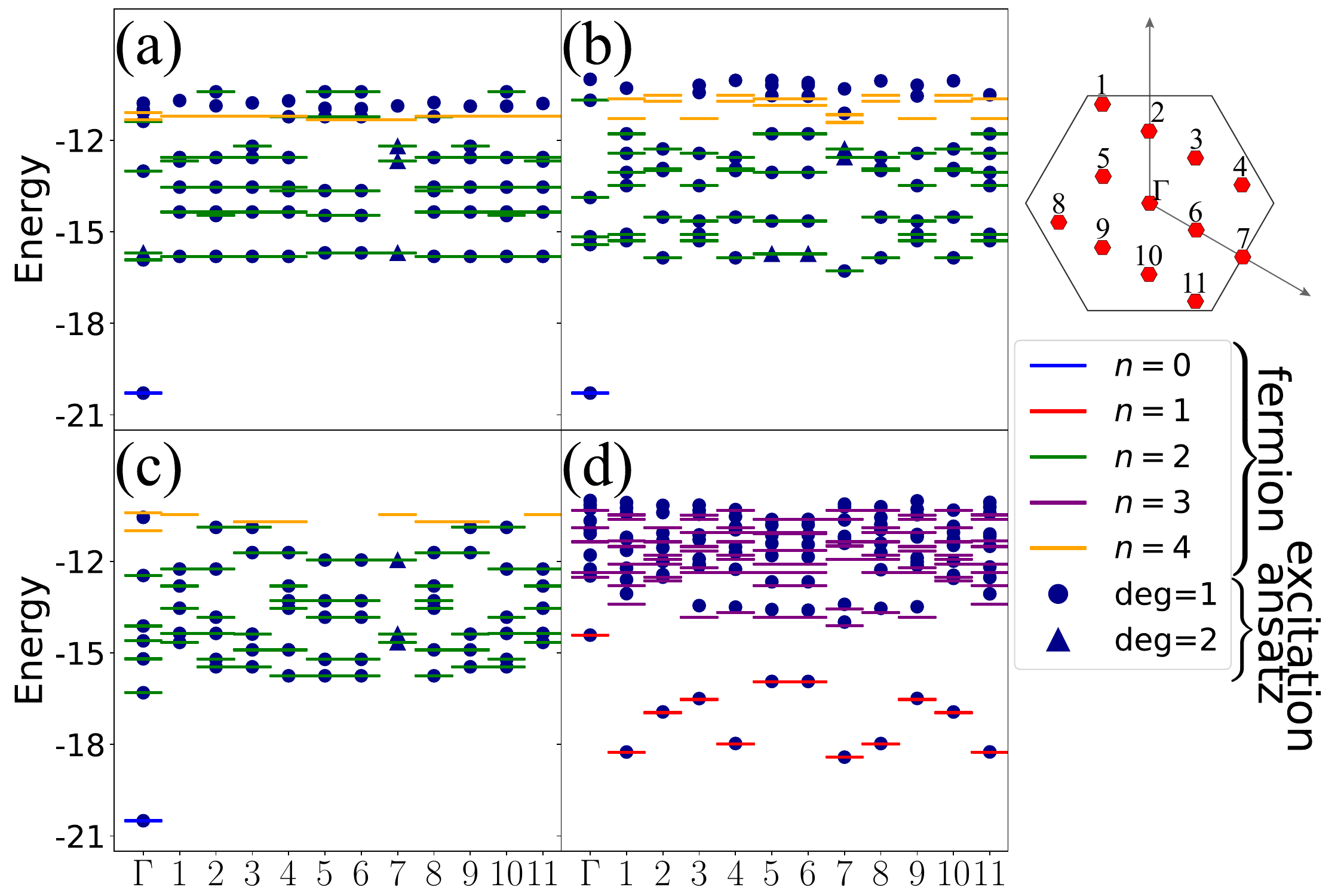}
    \caption{Zero vortex spectrum of the $24$-site torus. Subplot (a), (b), (c) and (d) correspond to sector $(\phi_1,\phi_2)=(+1,+1), (-1,+1), (+1,-1), (-1,-1)$, respectively.}
    \label{fig:Kitaev_zero_vortex_24site}
\end{figure}

Here we show additional variational excitation spectrum in zero vortex sectors. Using three ground states as in Sec.~\ref{sec:Kitaev_excitation}, the results for $(L_1,L_2)=(3,2)$ and $(4,3)$ are shown in Fig.~\ref{fig:Kitaev_zero_vortex_12site} and~\ref{fig:Kitaev_zero_vortex_24site}, respectively. In both cases, we also compare the results with fermion solution, so that finite size effects could be minimized and the numerical observations can be made general.

For both the $12$-site and $24$-site tori, the low energy spectrum with $0$-fermion and $2$-fermion states can be obtained with energy error smaller than $10^{-7}$, suggesting a scalability of our ansatz for this type of states. For $12$-site torus, due to the relatively small system size, the entire spectrum can be obtained within machine precision (error around $10^{-15}$), by using only single ground state. Note that the ground state should not be the exact ground state of the desired sector, otherwise the multi-fermion states would not be obtained. Instead, $U_1$ and $U_2$ play the role of fermion excitation operators, which give the high energy excitation subspace that can not be reached with only local two-site excitation operator.

For the $24$-site torus, the $1$-fermion states in $(-1,-1)$ sector can be obtained with energy error smaller than $4\times10^{-2}$, while for the $3$-fermion and $4$-fermion states, the error is significantly larger comparing to $N_s=12$. This indicates that the excitation ansatz has certain limitations in capturing multi-particle states (in the spin language).

\begin{figure}[htbp]
    \centering
    \includegraphics[width=0.95\columnwidth]{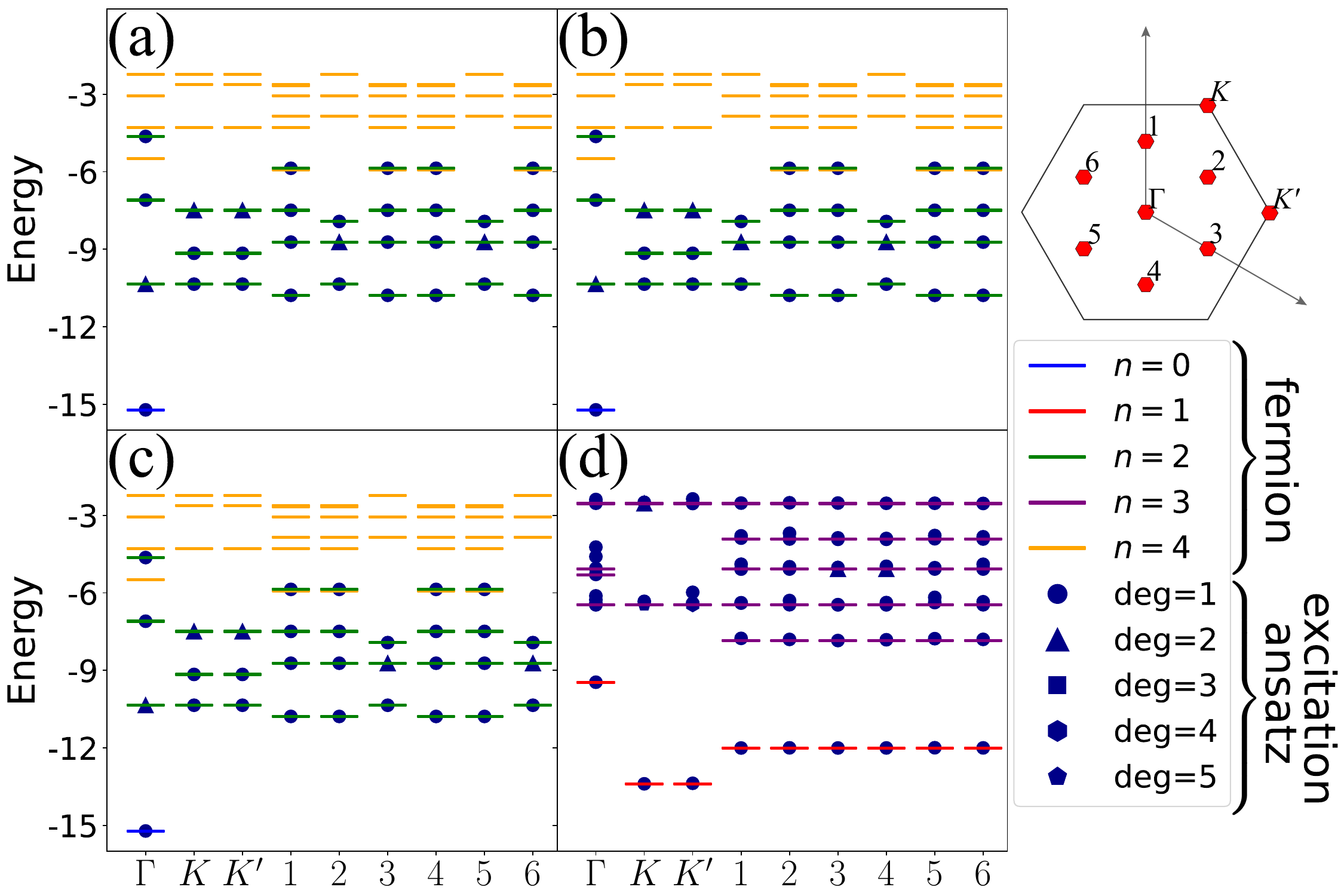}
    \caption{Zero vortex spectrum of the $18$-site torus with a single ground state. (a), (b), (c) and (d) correspond to sector $(\phi_1,\phi_2)=(+1,+1), (-1,+1), (+1,-1), (-1,-1)$, respectively.}
    \label{fig:Kitaev_zero_vortex_18site_1gs}
\end{figure}

\begin{figure}[htbp]
    \centering
    \includegraphics[width=0.95\columnwidth]{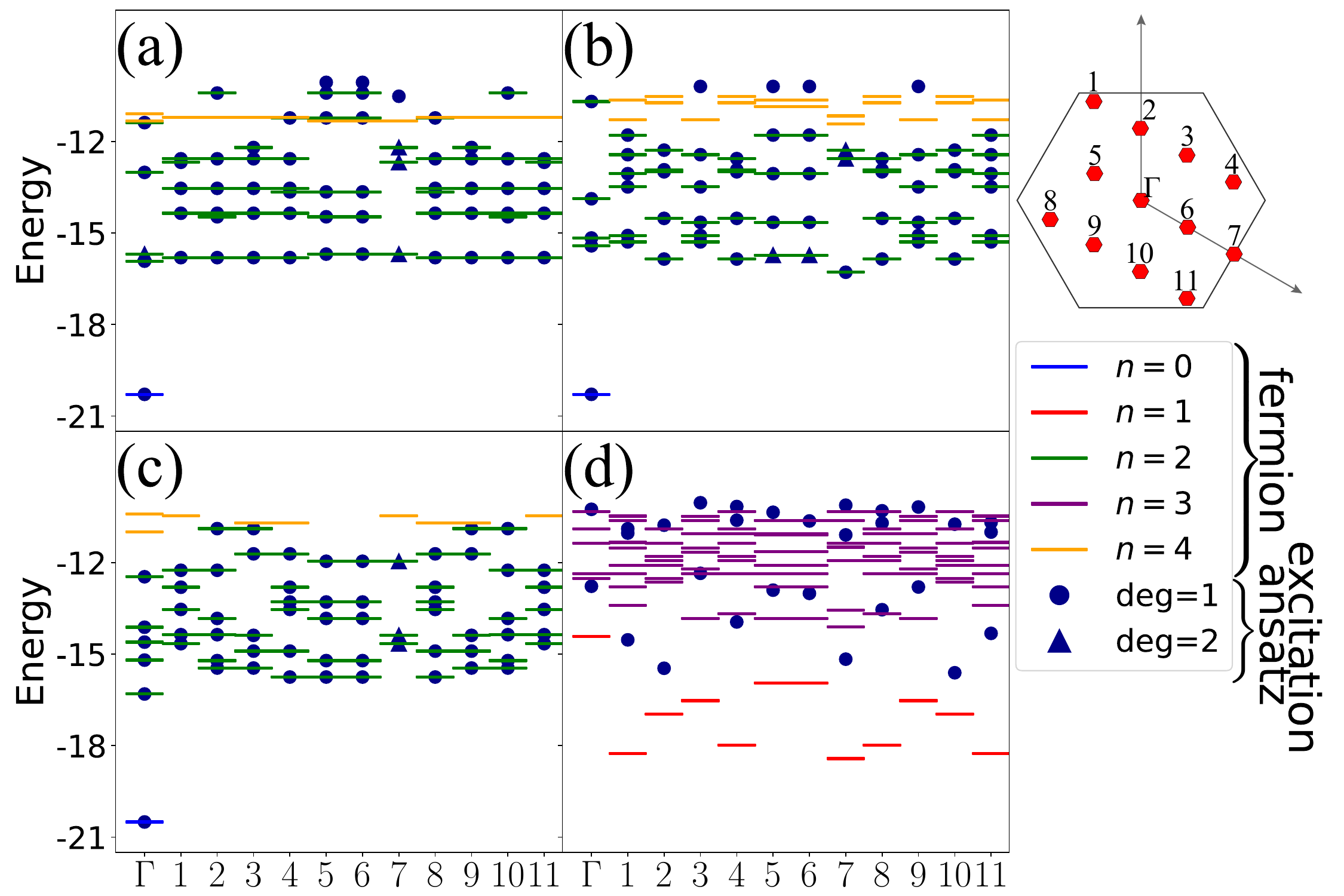}
    \caption{Zero vortex spectrum of the $24$-site torus with a single ground state. (a), (b), (c) and (d) correspond to sector $(\phi_1,\phi_2)=(+1,+1), (-1,+1), (+1,-1), (-1,-1)$, respectively.}
    \label{fig:Kitaev_zero_vortex_24site_1gs}
\end{figure}

As mentioned in Sec.~\ref{subsec:zero_vortex}, employing one ground state preparation circuit, we can obtain reasonably accurate low-energy spectrum with fewer basis states. Practically, for $(+1, +1)$, $(-1, +1)$ and $(+1, -1)$ sector, we choose the corresponding ground preparation circuit as the base for the insertion of excitation operators. While for $(-1, -1)$ sector, the circuit of $(+1, +1)$ sector is used and the excitation operators are inserted along with $U_{1}U_{2}$. On the $18$-site torus, for $0$-fermion and $2$-fermion states, this strategy achieves energy errors smaller than $2\times 10^{-9}$, and energy errors are smaller than $2\times 10^{-2}$ for $1$-fermion and $3$-fermion states (shown in Fig.~\ref{fig:Kitaev_zero_vortex_18site_1gs}). For the $24$-site torus, the maximum energy error of $0$-fermion and $2$-fermion states are similar to the results with three ground state circuits, while the $1$-fermion states can not be obtained within reasonable accuracy (see Fig.~\ref{fig:Kitaev_zero_vortex_24site_1gs}).

\subsection{Kitaev honeycomb model: Nonzero vortex sector}

\begin{figure}[h]
    \centering
    \includegraphics[width=0.95\columnwidth]{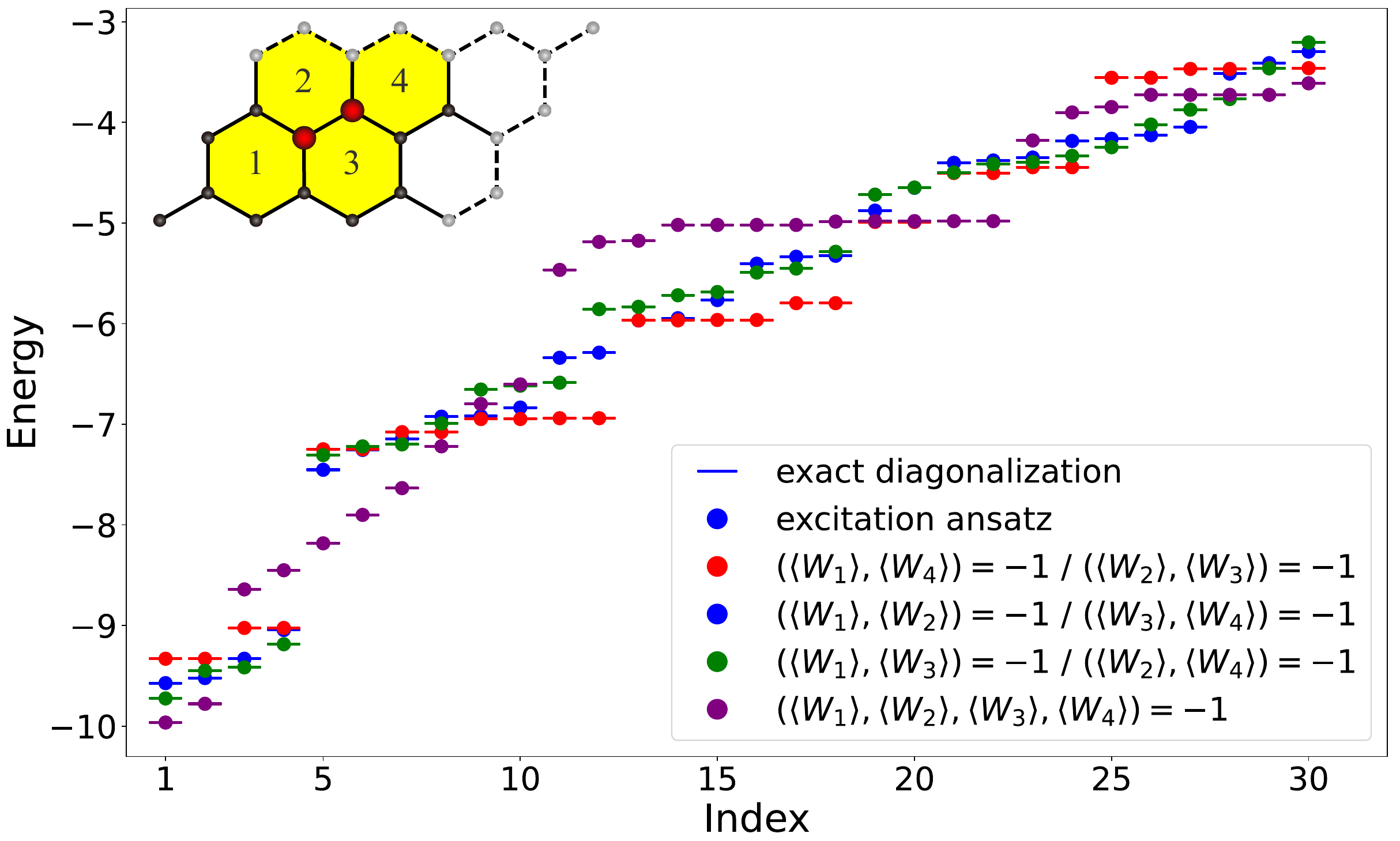}
    \caption{Nonzero vortex excitation spectrum of a $12$-site torus. The inset shows the label of vortex excitations adjacent to the highlighted bond.
    }
    \label{fig:Kitaev_nonzero_vortex_12site}
\end{figure}

For nonzero vortex sectors, besides the $18$-site results shown in the main text, we examine the results on a $12$-site torus. Similar to $18$-site, we focus on the vortex configurations around on specific bond, highlighted in inset of Fig.~\ref{fig:Kitaev_nonzero_vortex_12site}. As shown in Fig.~\ref{fig:Kitaev_nonzero_vortex_12site}, our ansatz achieves a high accuracy as in the zero vortex case, with the maximum energy error of $50$ lowest energy states being smaller than $10^{-10}$ for all $7$ nonzero vortex configurations.

\end{document}